\documentclass[11pt,american,twoside,a4paper]{article}

\usepackage[T1]{fontenc}
\usepackage[utf8]{inputenc}
\usepackage[american,provide=*]{babel}
\usepackage{a4wide}
\usepackage{fourier,bbm,bm}
\usepackage{amsmath,amsfonts,amsthm,amssymb}
\usepackage{color,xcolor,graphicx}
\usepackage{latexsym,mathtools,leftindex}
\usepackage[shortlabels]{enumitem}
\usepackage{xargs,xspace,needspace,xparse,easyReview}
\usepackage[active]{srcltx}
\usepackage[colorlinks=true]{hyperref}
\usepackage{fancyhdr}
\usepackage[toc]{appendix}
\usepackage{microtype}
\usepackage{subcaption} 
\usepackage[normalem]{ulem}
\usepackage{csquotes}
\usepackage[style=cap-alpha]{biblatex}
\usepackage{tikz}
\usetikzlibrary{calc}

\colorlet{darkblue}{blue!50!black}
\hypersetup{
	colorlinks,%
	citecolor=darkblue,%
	filecolor=red,%
	linkcolor=darkblue,%
	urlcolor=darkblue,%
	pdfnewwindow=true,%
	pdfstartview={FitH}
}

\numberwithin{equation}{section}
\renewcommand{\headheight}{14pt}

\NewDocumentCommand{\ASS}{mm}{\expandafter\newcommand\csname #1\endcsname{{\hyperref[#1]{\bf (#2)}}}}
\makeatletter
\newcommand{\nlabel}[2]{\begingroup
   \def\@currentlabel{{\bf (#2)}}%
   \label{#1}\endgroup
}
\makeatother

\newcommand{\assuming}[3]{
\begin{quote}\label{#2}{\bf(#1) }%
#3%
\end{quote}%
\ASS{#2}{#1}}
\newcommand{\ndef}[3]{%
\begin{definition}\label{#1}#3\end{definition}
\ASS{#1}{#2}
}
\theoremstyle{plain}
\newtheorem{theorem}{Theorem}[section]
\newtheorem{proposition}[theorem]{Proposition}
\newtheorem{lemma}[theorem]{Lemma}
\newtheorem{corollary}[theorem]{Corollary}
\theoremstyle{definition}
\newtheorem{definition}[theorem]{Definition}
\newtheorem{remark}{Remark}[section]
\newtheorem{remarks}[remark]{Remarks}
\newcommand{\bep}{\begin{proposition}}
\newcommand{\eep}{\end{proposition}}
\newcommand{\bel}{\begin{lemma}}
\newcommand{\eel}{\end{lemma}}
\newcommand{\bet}{\begin{theorem}}
\newcommand{\eet}{\end{theorem}}
\newcommand{\bed}{\begin{definition}}
\newcommand{\eed}{\end{definition}}
\newcommand{\bec}{\begin{corollary}}
\newcommand{\eec}{\end{corollary}}
\newcommand{\ber}{\begin{remark}}
\newcommand{\eer}{\end{remark}}
\newcommand{\beq}{\begin{equation}}
\newcommand{\eeq}{\end{equation}}
\newcommand{\bea}[1]{\begin{array}{#1}}
\newcommand{\eea}{\end{array}}
\def\proof{\noindent {\bf Proof.}\ \ }

\newcommand{\QED}{\hfill$\square$}
\newcommand{\ds}{\displaystyle}

\renewcommand{\i}{\mathrm{i}}

\newcommand{\e}{\mathrm{e}}
\renewcommand{\d}{\mathrm{d}}
\newcommand{\one}{\mathbbm{1}}
\renewcommand{\Im}{\operatorname{Im}}

\renewcommand{\sp}{\operatorname{sp}}
\newcommand{\spr}{\operatorname{spr}}
\newcommand{\spp}{\operatorname{\mathring{\mathrm{sp}}}}

\newcommand{\twol}[2]{\genfrac{}{}{0pt}{1}{#1}{#2}}
\newcommand{\bra}{\langle}
\newcommand{\ket}{\rangle}

\newcommand{\Ran}{\operatorname{Ran}}
\newcommand{\Ker}{\operatorname{Ker}}

\newcommand{\tr}{\operatorname{tr}}
\newcommand\ie{\textsl{i.e.,\ }}
\newcommand{\cA}{\mathcal{A}}
\newcommand{\cB}{\mathcal{B}}
\newcommand{\cC}{\mathcal{C}}

\newcommand{\cE}{\mathcal{E}}
\newcommand{\cF}{\mathcal{F}}
\newcommand{\cG}{\mathcal{G}}
\newcommand{\cH}{\mathcal{H}}

\newcommand{\cJ}{\mathcal{J}}
\newcommand{\cK}{\mathcal{K}}

\newcommand{\cN}{\mathcal{N}}
\newcommand{\cO}{\mathcal{O}}
\newcommand{\cP}{\mathcal{P}}

\newcommand{\cT}{\mathcal{T}}

\newcommand{\cV}{\mathcal{V}}
\newcommand{\cW}{\mathcal{W}}

\newcommand{\CC}{\mathbbm{C}}

\newcommand{\KK}{\mathbbm{K}}

\newcommand{\NN}{\mathbbm{N}}

\newcommand{\PP}{\mathbbm{P}}
\newcommand{\QQ}{\mathbbm{Q}}
\newcommand{\RR}{\mathbbm{R}}

\newcommand{\TT}{\mathbbm{T}}

\newcommand{\ZZ}{\mathbbm{Z}}

\newcommand{\fP}{\mathfrak{P}}

\newcommand{\fS}{\mathfrak{S}}

\newcommand{\bA}{\mathbf{A}}

\newcommand{\bsp}{{\boldsymbol{p}}}
\newcommand{\bsq}{{\boldsymbol{q}}}

\newcommand{\bsx}{{\boldsymbol{x}}}

\newcommand{\bseta}{{\boldsymbol{\eta}}}

\newcommand{\CPU}{\operatorname{CP_1}}
\newcommand{\BH}{\cB(\cH)}
\newcommand{\id}{\mathrm{id}}

\newcommand{\ep}{\operatorname{ep}}

\newcommand{\Ent}{\operatorname{Ent}}
\newcommand{\Ep}{\operatorname{Ep}}
\newcommand{\HS}{\mathrm{HS}}
\newcommand{\KMS}{\mathrm{KMS}}
\newcommand{\wP}{\widehat{\PP}}
\newcommand{\wJ}{\hat{\cJ}}

\newcommand{\ZZd}{{\ZZ_d}}

\newcommand{\doublearrow}[3][6pt]{%
  \draw[<->,thick,black!40!,shorten >=#1,shorten <=#1] (#2) -- (#3);%
}

\AtBeginEnvironment{theorem}{\Needspace{5\baselineskip}}
\AtBeginEnvironment{proposition}{\Needspace{3\baselineskip}}
\AtBeginEnvironment{definition}{\Needspace{5\baselineskip}}
\AtBeginEnvironment{corollary}{\Needspace{5\baselineskip}}
\AtBeginEnvironment{lemma}{\Needspace{5\baselineskip}}
\AtBeginEnvironment{quote}{\Needspace{5\baselineskip}}
\title{On entropy production of repeated quantum measurements~III. Quantum detailed balance}
\author{T. Benoist$^{1}$, N. Cuneo$^{2}$, V. Jak\v{s}i\'c$^{3}$, C-A. Pillet$^{4}$
\\ \\
$^1$Univ Toulouse, INUC, UT2J, INSA Toulouse, TSE, \\
CNRS, IMT, Toulouse, France.
\\ \\
$^2$Université Paris Cité and Sorbonne Université, CNRS,\\ Laboratoire de Probabilités, Statistique et Modélisation, F-75013 Paris, France
\\ \\
$^3$Dipartimento di Matematica, Politecnico di Milano, \\
piazza Leonardo da Vinci, 32, 20133 Milano, Italy 
\\ \\
$^4$Universit\'e de Toulon, CNRS, CPT, UMR 7332, 83957 La Garde, France\\
Aix-Marseille Univ, CNRS, CPT, UMR 7332, Case 907, 13288 Marseille, France
}
\begin{document}
\def\today{}
\maketitle

\centerline{\large \bf Dedicated to Joel  Lebowitz on the occasion of his 95th birthday}

\vspace{2em}

\noindent{\small{\bf Abstract.} In light of the dynamical-systems approach
to entropy production in repeated quantum measurements, proposed and illustrated in
\href{https://doi.org/10.1007/s00220-017-2947-1}{[CMP 357, 77--123 (2018)]}
and
\href{https://doi.org/10.1007/s10955-021-02725-1}{[JSP 182:44 (2021)]},
we characterize the KMS quantum detailed balance condition of quantum channels
by the time-reversal invariance and the vanishing of entropy production
of associated informationally complete quantum instruments.}
 
\tableofcontents 


\section{Introduction}
\label{sec:Intro}

This paper is a sequel to~\cite{Benoist2017c}, where a general approach to the
statistics of repeated quantum measurements was proposed, adopting the
philosophy of the thermodynamic formalism, and to~\cite{Benoist2021}, where this
new approach was illustrated by concrete examples. Our purpose here is
to revisit the concept of quantum detailed balance in light of these
previous works. In particular, we show that a quantum channel satisfies the KMS quantum detailed balance condition if and only if it can be associated to an informationally complete quantum instrument whose statistics has vanishing entropy production.

\subsection{Markov chains satisfying detailed balance}
\label{sec:DBMarkov}

In the classical world, the detailed balance condition is a characterization of
equilibrium based on time-reversal invariance and vanishing entropy
production. The notion can be traced back to the works of
Maxwell~\cite{Maxwell1867} and Boltzmann~\cite{Boltzmann1896}, and has a wide
range of applications. A historically celebrated one is Einstein's derivation
of Planck's blackbody radiation law~\cite{Einstein1916,Einstein1917}.
See~\cite[Section~8.2]{Aschbacher2006} for a pedagogical discussion of this
topic.

In the setting of a finite-state Markov chain $(x_n)_{n\in\NN}$ on the state space $\llbracket1,d\,\rrbracket=\{1,\ldots,d\}$, with transition matrix\footnote{$M_n(\KK)$
denotes the $\KK$-algebra of $n\times n$ matrices with entries in the field
$\KK$.} and  invariant probability row-vector
\begin{align*}
P&=[P_{ij}]\in M_d(\RR), & \pi&=[\pi_1,\ldots,\pi_d]\in\RR^d,\\[4pt]
P_{ij}&=\PP(x_{n+1}=j\,|\,x_n=i), &  \pi_i&=\PP(x_n=i)>0,
\end{align*}
the detailed balance condition takes the form
\beq
\label{db-mar}
\pi_i P_{ij}= \pi_j P_{ji}
\eeq
for all $i,j\in\llbracket1,d\,\rrbracket$. A practical route to the identification of
the quantum detailed balance condition starts with the observation that~\eqref{db-mar}
is equivalent to $P^\pi=P$, where $P^\pi$ denotes the adjoint of $P$ with respect
to the $\pi$-inner product on $\RR^d$ given by
\[
\bra x, y\ket_\pi=\sum_{i\in\llbracket1,d\,\rrbracket} \pi_i x_i y_i.
\]
A more general detailed balance condition is obtained by replacing~\eqref{db-mar} by the
condition $\widehat{P}=P$, where
\beq
\widehat{P}=\Theta^{-1}P^\pi\Theta,
\label{eq:ClassGenDB}
\eeq
and $\Theta\in M_d(\RR)$ is a permutation matrix such that
\beq
\pi\Theta=\pi,\qquad \Theta^2=I.
\label{eq:classinvol}
\eeq
This means that for some involutive permutation $\theta$ of the set $\llbracket1,d\,\rrbracket$ one has
$\pi_{\theta(i)}=\pi_i$ and 
\[\pi_{\theta(j)}P_{\theta(j)\theta(i)}=\pi_i P_{ij}\]
for all $i,j\in\llbracket1,d\,\rrbracket$. Note that the matrix $\Theta$ is
orthogonal w.r.t.\;both the standard and the $\pi$-inner product of $\RR^d$.
It is a simple exercise to show that Relation~\eqref{eq:ClassGenDB}
is equivalent to the time-reversal invariance of the Markov chain:
$\PP((x_1,\ldots,x_n)\in A)=\PP((\theta(x_n),\ldots,\theta(x_1))\in A)$
for each $n\in\NN$ and $A\subset\llbracket1,d\,\rrbracket^n$. This property
is itself equivalent to the vanishing of entropy production, as defined
below (see Section~\ref{sec:TRandEP}).

The more general condition~\eqref{eq:ClassGenDB} is of crucial relevance for
applications to statistical mechanics. The involution $\theta$ allows one to deal
with physical quantities, such as spin or momentum, which are odd (\ie
change sign) under time reversal. While it is often overlooked as an irrelevant
complication in discussions of detailed balance for Markov chains, its quantum
counterpart will be essential in establishing the equivalence of detailed
balance with the vanishing of entropy production for quantum channels.

\subsection{Quantum channels satisfying detailed balance}
\label{sec:DBQchannel}

In order to describe the quantum counterpart of the detailed balance
condition~\eqref{eq:ClassGenDB} let us introduce some  basic concepts and
notations. The $C^\ast$-algebra of all continuous linear operators on a Hilbert
space $\cH$ will be denoted by $\BH$, and its unit by $\one_\cH$ or simply $\one$.
The spectrum of a linear operator $X$ will be denoted by $\sp(X)$, its spectral
radius by $\spr(X)=\max|\sp(X)|$, and its peripheral spectrum by
$\spp(X)=\{\lambda\in\sp(X)\mid|\lambda|=\spr(X)\}$. The cone of positive
elements $X\geq 0$ of $\BH$ is written $\cB_+(\cH)$. The elements of $\BH$ are
{\sl observables} of the quantum system under consideration. A {\sl state} of
this system is a linear functional $\rho$ on $\BH$ taking non-negative values on
$\cB_+(\cH)$ and normalized by $\rho(\one)=1$. The number $\rho(X)$ is the
quantum expectation value of the observable $X$ when the system state is $\rho$.
A state is {\sl faithful,} written $\rho>0$, whenever $\rho(X^\ast X)=0$ implies
$X=0$. A linear transformation $\Phi$ of $\BH$ is {\sl positive} whenever it
maps $\cB_+(\cH)$ into itself. It is {\sl $n$-positive} whenever\footnote{We
denote by $\id$ the identity map on a $C^\ast$-algebra.} $\Phi\otimes\id$ is a
positive transformation of $\BH\otimes M_n(\CC)=\cB(\cH\otimes\CC^n)$, and
{\sl completely positive} if this holds for any $n\in\NN$. It is {\sl unital}
if it preserves the unit of $\BH$. A positive map on $\BH$ is {\sl irreducible}
whenever $\Phi(P)\le\lambda P$ for some orthogonal projection $P\in\BH$ and some
$\lambda>0$ implies $P\in\{0,\one\}$. We denote by $\CPU(\cH)$ the set of
completely positive unital maps on $\BH$ and note that, as a consequence
of~\cite[Corollary~1]{Russo1966}, $\spr(\Phi)=1$ for any $\Phi\in\CPU(\cH)$.

If $\cH$ is finite-dimensional, then $\BH$ carries a natural Hilbert space
structure with the Hilbert-Schmidt inner product $\bra X,Y\ket_\HS=\tr(X^\ast Y)$.
We denote by $\Phi^\ast$ the adjoint of a linear map $\Phi$ w.r.t.\;this duality.
Any state is then given by $X\mapsto\bra\rho,X\ket_\HS$, where $\rho$ is a
so-called {\sl density matrix,} a positive operator of unit trace. $\Phi$ is
positive/completely positive/irreducible iff $\Phi^\ast$ is. Moreover,
$\Phi$ is unital iff $\Phi^\ast$ is trace preserving so that, in particular,
it maps states into states.

If $\cH$ is infinite dimensional, then, as a Banach space, $\BH$ is the dual of
the Banach space $\cT(\cH)$ of all trace class operators on $\cH$, equipped with
the trace norm $\|T\|_1=\tr(\sqrt{T^\ast T})$. The duality is given by
$\cT(\cH)\times\BH\ni(T,X)\mapsto\bra T,X\ket=\tr(T^\ast X)$. The
corresponding weak*-topology on $\BH$ is also called ultraweak or
$\sigma\text{-weak}$. It is the weakest topology making the maps $\BH\ni
A\mapsto\bra T,A\ket$ continuous for all $T\in\cT(\cH)$. A {\sl normal
state} on $\BH$ is an ultraweakly continuous positive and normalized linear
functional. Such a state is induced by a density operator, \ie a positive
operator $\rho\in\cT(\cH)$ of unit trace through the formula\footnote{We will
identify density operators with the induced normal states.}
$$
\rho(X)=\bra\rho,X\ket=\tr(\rho X).
$$
If $\Phi_\ast$ is a CP-map on $\cT(\cH)$, then its dual is an ultraweakly
continuous CP-map $\Phi$ on $\BH$. It is unital whenever $\Phi_\ast$ is trace
preserving. Conversely, any ultraweakly continuous CP-map $\Phi$ on $\BH$
is dual to a CP-map on $\cT(\cH)$.

The extension of the notion of detailed balance to continuous quantum Markov semigroups has attracted
considerable attention in the literature, starting with the
works~\cite{Agarwal1973,Alicki1976,Carmichael1976,Kossakowski1977}; 
see~\cite[Section~1]{Fagnola2010} for a review of
the history of the subject. The literature on its discrete-time counterpart is scarcer; see~\cite{Dirac1924} for a pioneering work. However, 
it is the simplest setup for a quantum detailed balance condition, and  we focus on discrete-time dynamics.  Thus, in this work  the notion of quantum detailed balance will be  defined for quantum channels, \ie maps
$\Phi\in\CPU(\cH)$, and all quantum channels will act on a finite-dimensional Hilbert space. The continuous-time case will be considered elsewhere.\footnote{Taking an appropriate limit, many proofs in this work translate readily from discrete-time to continuous-time semigroups.}

By analogy with Section~\ref{sec:DBMarkov}, one proceeds as follows. Consider a
pair $(\Phi,\rho)$, where $\Phi\in\CPU(\cH)$ and the density matrix $\rho$ is
$\Phi$-invariant, $\Phi^\ast(\rho)=\rho$. Since $\Phi$ is unital, such a $\rho$ always exists. Moreover, one
can always restrict the analysis to the range of $\rho$, see~\cite[Proposition~9]{Baumgartner2012} or~\cite[Proposition~5.1]{Carbone2016},
and so without loss of generality we can assume that $\rho>0$.\footnote{If $\Phi$ is irreducible, then it has a unique and faithful invariant state.} In the following, a pair $(\Phi,\rho)$ always consists of a $\Phi\in\CPU(\cH)$
and a faithful $\Phi$-invariant state $\rho$.

The so-called KMS inner product on $\BH$ associated to $\rho$ is given by
$$
\bra X,Y\ket_\KMS=\tr(\rho^{\frac12}X^\ast\rho^{\frac12}Y),
$$
and we denote by $\Phi^\rho$ the adjoint of $\Phi$ w.r.t.\;this inner
product. One easily checks that
\[
\Phi^{\rho}: X\mapsto \rho^{-\frac12} \Phi^\ast(\rho^{\frac12} X\rho^{\frac12})\rho^{-\frac12},
\]
which implies that $\Phi^{\rho}(\one)=\one$ and $\Phi^{\rho\ast}(\rho)=\rho$.
It also follows that $\Phi^\rho$ is positive/completely positive/irreducible
iff $\Phi$ is. We note for later reference that if $X,Y\in\cB_+(\cH)$, then
\beq
\bra X,Y\ket_\KMS=\tr(Y^{\frac12}\rho^{\frac12}X\rho^{\frac12}Y^{\frac12})
\le\|X\|\tr(Y^{\frac12}\rho Y^{\frac12})
=\|X\|\bra\one,Y\ket_\KMS.
\label{eq:Holder}
\eeq

To simplify the exposition, we will call {\sl operator} any linear or
anti-linear map between complex vector spaces. To a unitary or anti-unitary operator $J$
on $\cH$, we associate the (anti-) $\ast$-morphism
$$
j:X\mapsto JXJ^\ast,
$$
which is unitary or anti-unitary w.r.t.\;the Hilbert-Schmidt inner product of
$\BH$. We note that $j$ is also unitary or anti-unitary w.r.t.\;the KMS inner
product whenever $J$ and $\rho$ commute, which can also be formulated as
$j(\rho)=\rho$. We will say that such an operator $J$ or the associated morphism
$j$ is $(\Phi,\rho)$-{\sl admissible} whenever
\beq
j(\rho)=\rho,\qquad\Phi(J^2)=\eta J^2,
\label{eq:Phi-rho-adm}
\eeq
for some phase $\eta\in\TT=\{z\in\CC\mid|z|=1\}$. A $(\Phi,\rho)$-admissible $J$ always
exists: the unit provides a unitary one, and the complex conjugation in an
eigenbasis of $\rho$ an anti-unitary one.

We set 
\beq
\widehat{\Phi}=j^{-1}\circ\Phi^{\rho}\circ j
\label{eq:defhatphi}
\eeq
as the quantum counterpart of~\eqref{eq:ClassGenDB}. Note that  $\widehat{\Phi}$  is positive/completely positive/irreducible iff
$\Phi$ is.  The following definition is
a non-commutative extension of the classical detailed balance condition
$\widehat{P}=P$, with the conditions~\eqref{eq:Phi-rho-adm} corresponding to the ones
in~\eqref{eq:classinvol}.
\ndef{QDB}{QDB}{
The pair $(\Phi,\rho)$ satisfies the {\bf Quantum Detailed Balance} condition, denoted
{\bf (QDB)}, if there exists a $(\Phi,\rho)$-admissible unitary or anti-unitary operator
$J$ on $\cH$ such that
$$
\widehat{\Phi}=\Phi.
$$
}

\begin{remarks}\label{rem:KMSpresQC} {\bf (i)} Since $\Phi(J^2)=J^2$ when $J^2=\one$,
the second condition in~\eqref{eq:Phi-rho-adm} should be understood as a
generalization of the involutive nature of $\Theta$.  We will show in Section~\ref{sec:value of c} that
this generalization is indeed necessary.

\noindent{\bf (ii)} This definition is closely related to the KMS inner product through
the definition of $\widehat{\Phi}$ in~\eqref{eq:defhatphi}. Similar detailed
balance conditions, based on different inner products on $\BH$ associated to the
faithful state $\rho$ have been introduced in the literature. They have various
motivations, a long history and intricate relationships;
see~\cite{Kossakowski1977,Fagnola2007, Fagnola2010,Amorim2021} and
references therein. In particular, the GNS quantum detailed balance condition is in general stronger than the KMS condition; see \cite[Proposition~7.1]{Fagnola2007}. Moreover, the adjoint of a completely
positive map w.r.t.\;the GNS inner product need not be completely positive; see the same proposition. We choose to work with the above definition because,
w.r.t.\;the KMS inner product, the adjoint of a completely positive map is again
completely positive. This will allow us to define the time reversal of
instruments modeling quantum measurements. This is not always possible for other
common notions of quantum detailed balance. The naturalness of the KMS inner product in the study of the
quantum detailed balance condition  in the continuous-time case  has been emphasized in~\cite[Section~4.4]{Derezinski2006}.

\noindent{\bf (iii)} We note that, according to our definition, $\Phi$
satisfies~\QDB{} whenever it is KMS-self-adjoint up to the conjugation,
$\Phi^\rho\circ j=j\circ\Phi$. 

\noindent{\bf (iv)} We do not assume that $J$ is an involution or even that
$J^2=\pm\one$.
\end{remarks}

\subsection{Repeated quantum measurements}
\label{sec:RepQMes}

Our main goal is to relate~\QDB{} to the characterization of equilibrium by
time-reversal invariance and vanishing entropy production.
Some results in this direction were obtained in~\cite{Fagnola2015}. Here we proceed by building on the recent
work~\cite{Benoist2017c,Benoist2021}. We shall relate the~\QDB{} condition for
an irreducible quantum channel $\Phi$ to the two properties of time-reversal
invariance and vanishing of entropy production of repeated quantum measurement
processes described by a suitable class of {\sl quantum instruments} associated
to $\Phi$.

\bed
Given a quantum channel $\Phi\in\CPU(\cH)$ and a Polish space $\bA$, a
$\boldsymbol{(\Phi,\bA)}${\bf-instrument} is a $\sigma$-additive map $\cJ$ from
the Borel $\sigma$-algebra $\cA$ of $\bA$ to the set of completely positive maps
on $\BH$ satisfying 
$$
\cJ(\bA)=\Phi.
$$
\eed

An instrument models a repeatable quantum measurement as follows. Let the system be in the
state $\rho$ at time $t=1$. A measurement is performed, and a random outcome
$\omega_1\in\bA$ is observed with the law $\tr(\cJ(\d\omega_1)^\ast(\rho))$.
After the measurement, the system state conditioned on $\omega_1\in A_1$ is
$$
\rho_{A_1}=\frac{\cJ(A_1)^\ast(\rho)}{\tr(\cJ(A_1)^\ast(\rho))}.
$$
The law of the outcome $\omega_2$ of the next measurement at time $t=2$ is
$\tr(\cJ(\d\omega_2)^\ast(\rho_{A_1}))$, and the state of the system  after the
second measurement conditioned on $(\omega_1,\omega_2)\in A_1\times A_2$ is
$$
\rho_{A_1A_2}=\frac{\cJ(A_2)^\ast(\rho_{A_1})}{\tr(\cJ(A_2)^\ast(\rho_{A_1}))}.
$$
The probability that the observed $(\omega_1,\omega_2)$ is  in $A_1\times A_2$ is
\[
\tr(\cJ(A_1)^\ast(\rho))\tr (\cJ(A_2)^\ast(\rho_{A_1}))
=\tr(\cJ(A_2)^\ast\circ\cJ(A_1)^\ast(\rho))
=\tr(\rho\cJ(A_1)\circ\cJ(A_2)(\one)).
\]
Continuing in this way, one derives that, after $n$ repeated measurements,
the probability of observing a sequence of outcomes
$(\omega_1, \ldots, \omega_n)\in A_1\times\dotsb\times A_n$
is given by
\beq
\PP_n(A_1\times\cdots\times A_n)=\tr\left(\rho\cJ(A_1)\circ\cdots\circ\cJ(A_n)(\one)\right).
\label{eq:Ppredef}
\eeq
This defines a probability measure $\PP_n$ on $\Omega_n=\bA^n$ and the unitality
of $\cJ(\bA)=\Phi$ implies that the family $(\PP_n)_{n\in\NN}$ is consistent.
Let $\PP$ be the unique probability measure induced on
$\Omega=\bA^\NN$, equipped with the usual product $\sigma$-algebra $\cF$, and
the filtration $(\cF_n)_{n\in\NN}$ generated by the cylinders
$$
[A_1,\ldots,A_n]=\{\omega\in\Omega\mid (\omega_1,\ldots,\omega_n)\in A_1\times\cdots\times A_n\},
\qquad
A_1,\ldots,A_n\in\cA.
$$
We shall rewrite \eqref{eq:Ppredef} as
\beq
\PP([A_1,\ldots,A_n])=\left\bra\one,\cJ(A_1)\cdots\cJ(A_n)\one\right\ket_\KMS.
\label{eq:Pdef}
\eeq
Let us denote by $\phi$ the left shift on $\Omega$. If the state $\rho$ is
$\Phi$-invariant, then, in terms of the KMS inner product on $\BH$, one has
\begin{align*}
\PP\circ\phi^{-1}([A_1,\ldots,A_n])
&=\PP_{n+1}(\bA\times A_1\times\cdots\times A_n)\\
&=\bra\one,\Phi\cJ(A_1)\cdots\cJ(A_n)\one\ket_\KMS\\
&=\bra\Phi^\rho\one,\cJ(A_1)\cdots\cJ(A_n)\one\ket_\KMS\\
&=\bra\one,\cJ(A_1)\cdots\cJ(A_n)\one\ket_\KMS\\
&=\PP([A_1,\ldots,A_n]).
\end{align*}
Thus, $\PP$ belongs to $\cP_\phi(\Omega)$, the set of $\phi$-invariant probability measures
on $(\Omega,\cF)$. The dynamical system $(\Omega,\phi,\PP)$ thus describes the
outcomes of our repeated measurement process. The measure $\PP$ will be called the
$\rho$-{\em statistics} of the instrument $\cJ$.

\subsection{Time reversal and entropy production}
\label{sec:TRandEP}

A {\em local reversal} on $\bA$ is a measurable involution $\theta:\bA\to\bA$.
The associated $\theta$-time reversal on $\Omega_n$ is the involution
\[
\theta_n(\omega_1,\ldots,\omega_n)=(\theta(\omega_n),\ldots,\theta(\omega_1)).
\]
If $\PP\in\cP_\phi(\Omega)$, then the family of probability measures
$(\wP_n)_{n\in\NN}$ defined by
$$
\wP_{n}=\PP_n\circ\theta_n
$$
is consistent. Hence, Kolmogorov's extension theorem yields a
unique $\wP\in\cP_\phi(\Omega)$ which describes the statistics of the
$\theta$-time reversal of the dynamical system $(\Omega,\phi,\PP)$.

We shall say that
the pair $(\Omega,\PP)$ is $\theta${\sl-time-reversal invariant}
if $\PP=\wP$, \ie  if $\PP_n=\wP_n$ for all $n\in\NN$.

We recall that the relative entropy of two probability measures $\PP$ and $\QQ$
on $(\Omega,\cF)$, defined by
$$
\Ent(\PP|\QQ)
=\begin{cases}
\ds\int_{\Omega}\log\frac{\d\PP}{\d\QQ}(\omega)\,\d\PP(\omega)&\text{if }\PP\ll\QQ;\\[16pt]
\infty&\text{otherwise,}
\end{cases}
$$
is non-negative and vanishes iff $\PP=\QQ$. We define the $\theta$-{\sl entropy
production} of $(\Omega,\phi,\PP)$ in the discrete-time interval
$\llbracket1,n\rrbracket$ by
\[
\Ep(\PP_n,\theta)=\Ent(\PP_n|\wP_n).
\]
The (possibly infinite) non-negative number
\[
\ep(\PP,\theta)=\limsup_{n\to\infty}\tfrac1n\Ep(\PP_n,\theta)
\]
is called the $\theta$-{\sl entropy production rate} of $(\Omega,\phi,\PP)$.
For further discussion of this important  notion we refer the
reader to~\cite{Benoist2017c}.

At the current level of generality, the $\theta$-entropy production rate can
exhibit pathological behavior. In~\cite{Andrieux2016} one can find a striking example
of a $\phi$-ergodic $\PP$ such that, for any local reversal $\theta$, $\PP$ and
$\wP$ are mutually singular while $\ep(\PP,\theta)=0$. The following
upper-decoupling property (using the terminology of \cite{Cuneo2019}) precludes these pathologies, ensuring that
$\theta$-time-reversal invariance is equivalent to the vanishing of entropy
production, and in this sense is characteristic of equilibrium.

\assuming{UD}{ER}{There is a constant $C>0$ such that for any $A\in\cF_n$
with $n\in\NN$ and $B\in\cF$,
$$
\PP(A\cap\phi^{-n}(B))\leq C\,\PP(A)\PP(B).
$$
}

Note that if \ER{} holds for $\PP$, then it also holds for $\wP$. The following
result is an immediate generalization of the proofs of Theorem 2.1 and
Proposition 2.2 in~\cite{Benoist2017c} using the Donsker--Varadhan variational
formula for the relative entropy. For the reader's
 convenience, we provide the
proof in Section~\ref{sec:proof-TRI-ep}.

\bep \label{prop:conv_ep} Suppose that~\ER{}  holds. Then:
\begin{enumerate}[label=(\roman*)]
\item The following (possibly infinite) limit exists:
\[
\ep(\PP,\theta)=\lim_{n\to\infty}\tfrac1n\Ep(\PP_n,\theta).
\]
\item Assume in addition that $\PP$ is $\phi$-ergodic. Then $\ep(\PP,\theta)=0$
iff $(\Omega,\PP)$ is $\theta$-time-reversal invariant.
\end{enumerate}
\eep

Let $\cJ$ be a $(\Phi,\bA)$-instrument, and $(\Omega,\PP)$ its
$\rho$-statistics. Given a local reversal $\theta$, a $\theta$-time reversal of
the pair $(\cJ,\rho)$ is any pair of quantum instrument and state
$(\wJ,\hat{\rho})$ such that the $\hat\rho$-statistics of $\wJ$ is
$(\Omega,\wP)$. In these circumstances, we shall write $\ep(\cJ,\rho,\theta)$ for
$\ep(\PP,\theta)$.

Going back to~\cite{Crooks2008}, a canonical choice for this $\theta$-time reversal
is given by
\beq
\wJ(A)=j^{-1}\circ\cJ(\theta(A))^\rho\circ j,\qquad\hat\rho=\rho,
\label{Crooks}
\eeq
where $j$ is any $(\Phi,\rho)$-admissible (anti-)unitary map on $\cB(\cH)$.
Indeed, using the fact that $j^\rho=j^{-1}$ and $j(\one)=\one$ gives, in the unitary case,
\begin{align}
\bra\one,\wJ(A_1)\cdots\wJ(A_n)\one\ket_\KMS
&=\bra\one,j^\rho\cJ(\theta(A_1))^\rho\cdots\cJ(\theta(A_n))^\rho j\one\ket_\KMS\nonumber\\
&=\bra\cJ(\theta(A_n))\cdots\cJ(\theta(A_1))\one,\one\ket_\KMS\nonumber\\
&=\overline{\bra\one,\cJ(\theta(A_n))\cdots\cJ(\theta(A_1))\one\ket_\KMS}\label{eq:hatJform}\\
&=\overline{\PP([\theta(A_n),\ldots,\theta(A_1)])}\nonumber\\
&=\wP([A_1,\ldots,A_n]).\nonumber
\end{align}
A similar calculation with the same result holds in the anti-unitary case.

\subsection{Dilations of channels and instruments}
\label{sec:instruments POVM}

Instruments are related to the more commonly known notion of positive
operator-valued measure.

\bed
Let $\cE$ be a separable Hilbert space and $(\bA,\cA)$ a measurable space.
A map
$$
M:\cA\to\cB_+(\cE)
$$
is a {\bf Positive Operator-Valued Measure} (POVM) if $M(\bA)=\one$ and,
for all $x,y\in\cE$, the set-function $\cA\ni A\mapsto\bra x,M(A)y\ket$
is $\sigma$-additive.
\eed

From this definition it follows that, for any POVM $M:\cA\to\cB_+(\cE)$ and any
state $\gamma$ on $\cB(\cE)$, the map $A\mapsto\gamma\circ M(A)$ defines a
probability measure on $\cA$. It is interpreted as the probability to obtain a
measurement result $a\in A$ when the system is in the state $\gamma$. However,
a POVM does not provide a way to determine the state of the system after the
measurement. This is the primary role of a $(\Phi,\bA)$-instrument, as described
in Section~\ref{sec:RepQMes}. The following results relate channels and their related
instruments to POVMs. The proof is given in Section~\ref{sec:proofDilation}.

\bep\label{prop:dilation}
Let $\cH$ be a Hilbert space of finite dimension $d$.
\begin{enumerate}[label=(\roman*)]
\item A linear map $\Phi:\BH\to\BH$ is completely positive iff there exists a
Hilbert space $\cE$ and a linear map $V:\cH\to\cH\otimes\cE$ such that,
for any $X\in\BH$,
\begin{align}
\Phi(X)=V^\ast(X\otimes\one_{\cE})V.
\label{eq:Stinespring dilation}
\end{align}
Moreover, $\Phi$ is unital iff $V$ is isometric. One calls such a pair $(\cE,V)$
a Stinespring dilation of $\Phi$. This dilation is said to be minimal whenever the linear span of
$(\BH\otimes\one_\cE)V\cH$ is dense in $\cH\otimes\cE$. A minimal dilation
always exists, with $\dim(\cE)\le d^2$, and is unique up to unitary equivalence.
\item Let $(\cE,V)$ be a Stinespring dilation of $\Phi\in\CPU(\cH)$, and $\cJ$ be a
$(\Phi,\bA)$-instrument. Then, there exists a POVM $M:\cA\to\cB_+(\cE)$ such that
$$
\cJ(A)(X)=V^\ast(X\otimes M(A))V
$$
for any $A\in\cA$ and $X\in\BH$. We shall say that $M$ is a $({\cJ},\cE,V)$-POVM.
\end{enumerate}
\eep

Formula~\eqref{eq:Stinespring dilation} is a special form of Stinespring's
dilation~\cite{Stinespring1955}, which holds more generally for arbitrary Hilbert spaces
$\cH$, provided the CP map $\Phi$ is ultraweakly continuous.

As mentioned in Proposition~\ref{prop:dilation}(i), for finite-dimensional $\cH$,
one can always choose $\cE$ to be finite dimensional too. However, for possible
future reference, dilation spaces $\cE$ will only be assumed to be separable and may
be infinite-dimensional.

\subsection{Main result}
\label{sec:mainResult}

A POVM $M$ defines a map $\gamma\mapsto\gamma\circ M$ from normal states $\gamma$
on $\cB(\cE)$ to probability measures on $\cA$. If this map is injective, then $M$ allows one to
identify the state $\gamma$ from the statistics of outcomes of measurements of $M$ on independent
copies of $\gamma$. The associated instruments are central to our results
as they allow for the interpretation of the measure $\PP$ as a purely generated finitely
correlated state; see Section~\ref{sec:TRI implies QDB}.

\bed
Let $\cE$ be a separable Hilbert space, $\cN$ the set of normal states on $\cB(\cE)$,
and $(\bA,\cA)$ a measurable space.
\begin{enumerate}[label=(\roman*)]
\item A POVM $M:\cA\to\cB_+(\cE)$ is called {\bf Informationally Complete} (IC-POVM)
if the map
$$
\cN\ni\gamma\mapsto\gamma\circ M
$$
is injective.
\item Given a $(\Phi,\bA)$-instrument $\cJ$, a local reversal $\theta$ is
unitarily (resp.~anti-unitarily) {\bf implementable} if there exists a Stinespring dilation $(\cE,V)$ of $\Phi$,
a $(\cJ,\cE,V)$-POVM $M$, and a unitary (resp.~anti-unitary) operator $\Theta$ on
$\cE$ such that
$$
M\circ\theta(\,\cdot\,)=\Theta^\ast M(\,\cdot\,)\Theta.
$$
\item A $(\Phi,\bA)$-instrument  $\cJ$ is called {\bf informationally complete} if there
exists a Stinespring dilation $(\cE,V)$ of $\Phi$ and an associated $(\cJ,\cE,V)$-POVM
that is informationally complete.
\item Given an informationally complete $(\Phi,\bA)$-instrument $\cJ$, a local reversal $\theta$ is unitarily (resp.~anti-unitarily) {\bf IC-implementable} if the POVM $M$ in (ii) can be chosen informationally complete.
\end{enumerate}
\eed

In~\cite{Crooks2008}, the maps $\theta$ and $j$ involved in Relation~\eqref{Crooks}
were both the identity map, which is insufficient for our purposes. In particular,
regarding the choice of reversal $\theta$, implementability will play an important role
in our analysis.

\begin{remarks} 
{\bf (i)}
If the informationally complete instrument $\cJ$ admits an IC-implementable local reversal
$\theta$, then the operator $\Theta$ can be chosen such that $\Theta^2$ is involutive.
Indeed, it follows from
$$
M(A)=M\circ\theta^2(A)=\Theta^{\ast2}M(A)\Theta^2
$$
that $[\Theta^2,M(A)]=0$ for every $A\in\cA$. $M$ being informationally complete, Lemma~\ref{lem:IC}(ii)
allows us to conclude that there is $\varphi\in\RR$ such that $\Theta^2=\e^{\i\varphi}\one_\cE$.
If $\Theta$ is unitary, the phase factor $\e^{-\i\varphi/2}$ can be absorbed in the definition of
$\Theta$ so that it becomes an involution. If $\Theta$ is anti-unitary, it follows from Wigner's
decomposition theorem~\cite{Wigner1993a} that $\Theta^2=\pm\one_\cE$, so $\Theta^2$ is involutive.

\noindent{\bf (ii)} Informationally complete POVMs and instruments are of considerable theoretical and experimental importance with 
a large literature devoted to them. For an introduction to the topic that is in the spirit of our work, we refer  the reader to~\cite{Busch2016}. To the best of our knowledge, they have not been studied before in the context of the quantum detailed balance condition. 

\end{remarks}

We are now in a position to formulate our main result, which relates the quantum detailed
balance condition to the vanishing of the entropy production rate for irreducible
quantum channels with a faithful invariant state.

\bet\label{thm:main}
If\, $\Phi\in\CPU(\cH)$ is irreducible, then it satisfies~\QDB{} iff there exists an
informationally complete $(\Phi,\bA)$-instrument $\cJ$ and an IC-implementable
local reversal $\theta$ such that  $\ep(\cJ,\rho,\theta)=0$.
\eet

The remaining parts of this paper are organized as follows.

The proof of Theorem~\ref{thm:main} is given in Section~\ref{sec:submain}, 
where we introduce our main tool: a lift of the~\QDB{} condition to the instrumental level. 
We will derive Theorem~\ref{thm:main} from two results,
Theorems~\ref{thm:main-exist-instrument} and~\ref{main-instrument-QDB}, both of
independent interest. Instrumental detailed balance will be related to
the~\QDB{} condition by the first one, and to time-reversal invariance and vanishing
entropy production rate by the second one.
In Section~\ref{sec:value of c}, we discuss structural constraints
on the choice of $J$ and the associated values of $\eta$. 
We show in particular that we cannot avoid considering anti-unitary $J$, $\eta\neq1$ and $J^2\neq\one$.
Sections~\ref{sec:proof-TRI-ep}--\ref{sec:proof possible values of c} are devoted to the
proofs of Propositions~\ref{prop:conv_ep}, \ref{prop:dilation}, \ref{prop:prop channel}
and~\ref{prop:possible values of c}. 
In Sections~\ref{sec:prelimChannels}--\ref{sec:prelimPOVM} we state and prove some
preliminary lemmas on channels, instruments and POVMs. In
Section~\ref{sec:Stinespring}, we elaborate on the relations between Stinespring
dilations of $\Phi$ and its reversal $\widehat{\Phi}$. These relations lead, in
Section~\ref{sec:QDB implies IQDB}, to the proof of
Theorem~\ref{thm:main-exist-instrument}. In Section~\ref{sec:PGFCS}, we
introduce another central tool, purely generated finitely correlated states. We
prove a slight extension of a result from~\cite{Fannes1994}, essentially
following the alternative proof of~\cite{Guta2015}.
This result is used in Section~\ref{sec:TRI implies QDB} to prove
Theorem~\ref{main-instrument-QDB}.

\paragraph*{Acknowledgments.} The work of CAP and VJ was partly funded by the CY Initiative grant Investisse\-ments d’Avenir, grant number ANR-16-IDEX-0008. VJ acknowledges the support of NSERC and the support of the MUR
grant "Dipartimento di Eccellenza 2023-2027" of Dipartimento di Matematica, Politecnico di Milano.
We also acknowledge the support of the ANR project DYNACQUS, grant number ANR-24-CE40-5714. We acknowledge using Large Language Models to find typos and oversights in the text. The required corrections were implemented by hand at the same time as the remarks by the referees.

\vspace{-1.5em}

\paragraph{Data availability.} No data was used or produced for the redaction of this article.

\vspace{-1.5em}

\paragraph{Conflict of interest.} The authors do not have any conflict of interest related to this work.

\section{Instrumental detailed balance}
\label{sec:submain}

\ndef{IQDB}{IQDB}{A $(\Phi,\bA)$-quantum instrument $\cJ$ is said to satisfy
the {\sl instrumental quantum detailed balance condition} {\bf(IQDB)} if there exists
a $(\Phi,\rho)$-admissible operator $J$ on $\cH$ and a local reversal $\theta$ on $\bA$, such that
\[
\cJ=\wJ,
\]
where $\wJ$ is the reversed instrument defined in \eqref{Crooks}.
}

\begin{remarks}
{\bf (i)} The implication  (i)$\implies$(ii) in Theorem~\ref{main-instrument-QDB} shows that if~\IQDB{} holds, then $\theta$ is unitarily (resp. anti-unitarily) implementable when $J$ is anti-unitary (resp. unitary).

\noindent{\bf (ii)} Since $\theta(\bA)=\bA$ and $\cJ(\bA)=\Phi$, comparing~\eqref{eq:defhatphi}
and~\eqref{Crooks} yields that the~\IQDB{} condition implies the~\QDB{} condition.
Moreover, under~\IQDB, $\wP=\PP$ follows from~\eqref{eq:hatJform} so that,
consequently, $\ep(\cJ,\rho,\theta)=0$.
\end{remarks}

The following result relates conditions~\IQDB{} and~\QDB.

\bet \label{thm:main-exist-instrument} A pair $(\Phi,\rho)$ satisfies~\QDB{} if and only if there exists a Polish space $\bA$ and an informationally complete $(\Phi,\bA)$-instrument satisfying~\IQDB{} with an IC-implementable
local reversal.
\eet
Note that this statement is trivial if the instrument is not required to be informationally complete.
Our proof will show that the same operator $J$ can be used in
conditions~\QDB{} and~\IQDB{}.

The next  result links~\IQDB{} to central physical properties: time-reversal invariance and
vanishing of the entropy production rate.

\bet\label{main-instrument-QDB}
Let the pair $(\Phi,\rho)$ be equipped with a $(\Phi,\bA)$-instrument $\cJ$ and a local reversal $\theta$.
Let $\PP$ be the $\rho$-statistics of $\cJ$, $\wP$ its $\theta$-time reversal,
and $\ep(\cJ,\rho,\theta)$ the associated $\theta$-entropy production rate.
Consider the following statements:
\begin{enumerate}[label=(\roman*)]
\item $\cJ$ satisfies an~\IQDB{} condition with local reversal $\theta$ and anti-unitary (resp. unitary)
$(\Phi,\rho)$-admissible $J$.

\item $\theta$ is unitarily (resp. anti-unitarily) implementable and $\wP=\PP$.
\item $\theta$ is unitarily (resp. anti-unitarily) implementable and $\ep(\cJ,\rho,\theta)=0$.
\end{enumerate}
Then (i)$\implies$(ii)$\implies$(iii). Moreover, if $\Phi$ is irreducible then
(iii)$\implies$(ii), and if additionally $\cJ$ is informationally
complete and $\theta$ is unitarily (resp.~anti-unitarily) IC-implementable, then (ii)$\implies$(i).
\eet
\begin{remark} It is not difficult to construct an  example on ${\cal H}=\CC^3$ for which the implication (ii)$\implies$(i) fails if the assumption 
that $\cJ$ is informationally complete is omitted.  
\end{remark}
{\noindent\bf Proof of Theorem~\ref{thm:main}}. By Theorem~\ref{thm:main-exist-instrument},
the~\QDB{} condition is equivalent to the~\IQDB{} condition with an IC-implementable local reversal $\theta$.
The equivalence (iii)$\Longleftrightarrow$(i) of Theorem~\ref{main-instrument-QDB} completes
the proof.
\QED

\section{On the choice of $J$ and the possible values of $\eta$} 
\label{sec:value of c}

Given a pair $(\Phi,\rho)$ satisfying the \QDB{} condition, there may exist
several $(\Phi,\rho)$-admissible $J$'s such that $j^{-1}\circ\Phi^\rho\circ
j=\Phi$, with possibly distinct values of $\eta$ in~\eqref{eq:Phi-rho-adm}.
In this section we show that there are some constraints on the possible choices.
In particular, we show that there may not be a choice of $J$ such that $\eta=1$, in
which case it is impossible to choose $J$ as an involution, and that we cannot
avoid considering anti-unitary $J$.
 
In order to simplify the discussion, we restrict ourselves to irreducible channels
$\Phi$. Like for the transition matrix of a Markov chain, irreducibility has
important consequences on the spectral and ergodic properties of channels. 
The following result is well-known~\cite{Evans1987, Kuemmerer2003}; see Section~\ref{sec:proof prop channel} for a proof.

\bep\label{prop:prop channel}
Assume that $\Phi\in\CPU(\cH)$ is irreducible. Then,
\begin{enumerate}[label=(\roman*)]
\item
$\spp(\Phi)$ is a finite subgroup of the unit circle, \ie
there exists an integer $p>0$, the \textbf{period} of $\Phi$, such that 
$$
\spp(\Phi)=\TT_p=\{\xi_p^a\mid a\in\llbracket0,p-1\rrbracket\},\qquad\xi_p=\e^{2\pi\i/p}.
$$
Moreover, each peripheral eigenvalue is simple.
\item There exists a unique (up to labeling) orthogonal partition of unity
$$
\one=\bigoplus_{\alpha\in\TT_p}P_\alpha,
$$ 
such that $\Phi(P_\alpha)=P_{\tau(\alpha)}$, where $\tau$ denotes the
group translation $\tau(\alpha)=\xi_p^{-1}\alpha$. 
Such a partition is called a \textbf{maximal cycle} of $\Phi$.
\item The unitary operator 
$$
U=\sum_{\alpha\in\TT_p}\alpha P_\alpha
$$
is such that $\Phi(U^nX)=\xi_p^nU^n\Phi(X)$ for all $X\in\cB(\cH)$ and $n\in\ZZ$.
\item There exists a unique density matrix $\rho$ such that $\Phi^\ast(\rho)=\rho$. Moreover, 
$\rho >0$ and
$$
\rho=\bigoplus_{\alpha\in\TT_p}P_\alpha\rho P_\alpha.
$$
\item For every $(\Phi,\bA)$-instrument, the associated $\rho$-statistics $\PP$ is $\phi$-ergodic.
\end{enumerate}
\eep

Since the faithful $\Phi$-invariant state $\rho$ is unique, in the remaining parts of this section 
we shall omit its explicit mention when possible. 

\subsection{Non-uniqueness of $\eta$ and some forbidden values}
\label{sec:non unique c}

First, we determine the possible values of $\eta$ depending on the
period of $\Phi$ and the nature of $J$. 

The following proposition is
proved in Section~\ref{sec:proof possible values of c}.

\begin{proposition}\label{prop:possible values of c}
Assume that $\Phi$ is irreducible of period $p$ and satisfies \QDB. Let $J$ be a unitary 
or anti-unitary operator such that $j^{-1}\circ\Phi^{\rho}\circ j=\Phi$,
and $\Phi(J^2)=\eta J^2$ for some $\eta\in\TT_p$. Denote by $\widetilde\TT_p$
the subgroup generated by $\xi_p^2$.

\begin{enumerate}[label=(\roman*)]
\item\label{it:c pm1} If $J$ is unitary, then $\eta\in\{-1,+1\}$. 
\item\label{it:possible values of c} If $J$ is anti-unitary, there exists a 
family $(J_\alpha)_{\alpha\in\widetilde\TT_p}$ of anti-unitary operators such that 
$$
j_\alpha^{-1}\circ\Phi^{\rho}\circ j_\alpha=\Phi,\qquad\Phi(J_\alpha^2)=\eta\alpha J_\alpha^2.
$$
In particular, if $p$ is odd, then $\widetilde\TT_p=\TT_p$, and any $p$-th root of unity is a possible
value of $\eta$.
\end{enumerate}
\end{proposition}

Second, we establish that some channels have constraints on the choice of $J$ and the associated 
values of $\eta$. All the remaining propositions of this section are proved in Section~\ref{sec:xpl value of c}.

\begin{proposition}\label{prop: no unitary no 1}
\begin{enumerate}[label=(\roman*)]
\item For any root of unity $\eta$, there exists an irreducible quantum channel $\Phi$ satisfying \QDB{} with 
anti-unitary $J$ such that $\Phi(J^2)=\eta J^2$.
\item For any even integer $p\geq4$, there exists an irreducible quantum channel $\Phi$ satisfying \QDB{} 
with anti-unitary $J$ such that $\Phi(J^2)=\xi_pJ^2$ but there does not exist an anti-unitary operator $J$ 
for which $\Phi$ satisfies \QDB{} with $\Phi(J^2)=J^2$.
\end{enumerate}
\end{proposition}

We then obtain the following corollary.

\begin{corollary}\label{cor:noone}
There exists an irreducible quantum channel $\Phi$ satisfying \QDB{} with an anti-unitary $J$ such 
that $\Phi(J^2)=-J^2$, but not satisfying \QDB{} with any anti-unitary $J$ such that $\Phi(J^2)=J^2$.
\end{corollary}

\proof
Taking $p=6$ in Proposition~\ref{prop: no unitary no 1}(ii) gives an irreducible channel $\Phi$ satisfying the second assertion of the corollary and an anti-unitary $\widetilde{J}$ such that $\Phi(\widetilde{J}^2)=\xi_6\widetilde{J}^2$. Since $\xi_6$ is a peripheral eigenvalue of $\Phi$, if $q$ denotes its period then $\alpha=\xi_6^2\in\widetilde\TT_q$. With this $\alpha$, the anti-unitary $J=\widetilde{J}_\alpha$ given by Proposition~\ref{prop:possible values of c}(ii) satisfies $\Phi(J^2)=\xi_6\alpha J^2=\xi_6^3J^2=-J^2$.
\QED

The next result shows that we cannot dispense with considering anti-unitary $J$.
\begin{proposition}\label{prop:anti U identity nu unitary}
There exists an irreducible quantum channel $\Phi$ satisfying \QDB{} with an anti-unitary $J$ such that $J^2=\one$, but not satisfying
\QDB{} for any unitary $J$.
\end{proposition}

We turn to a result similar to Corollary~\ref{cor:noone}, but for unitary $J$.
\begin{proposition}\label{prop:unitary no 1}
There exists an irreducible quantum channel $\Phi$ satisfying \QDB{} with a unitary $J$ such that $\Phi(J^2)=-J^2$, but not satisfying \QDB{} with any unitary $J$ such that $\Phi(J^2)=J^2$.
\end{proposition}

\begin{remark}
Proposition~\ref{prop:anti U identity nu unitary} shows that there exists a quantum channel satisfying \QDB{} with some anti-unitary $J$ but no unitary $J$. The converse question remains open: we do not know whether there exists a quantum channel satisfying \QDB{} with some unitary $J$  but no anti-unitary $J$.
\end{remark}

\subsection{A class of channels illustrating various constraints}
\label{sec:xpl value of c}

The proof of Propositions~\ref{prop: no unitary no 1} to~\ref{prop:unitary no 1}
is based on a class of channels which we now introduce. 

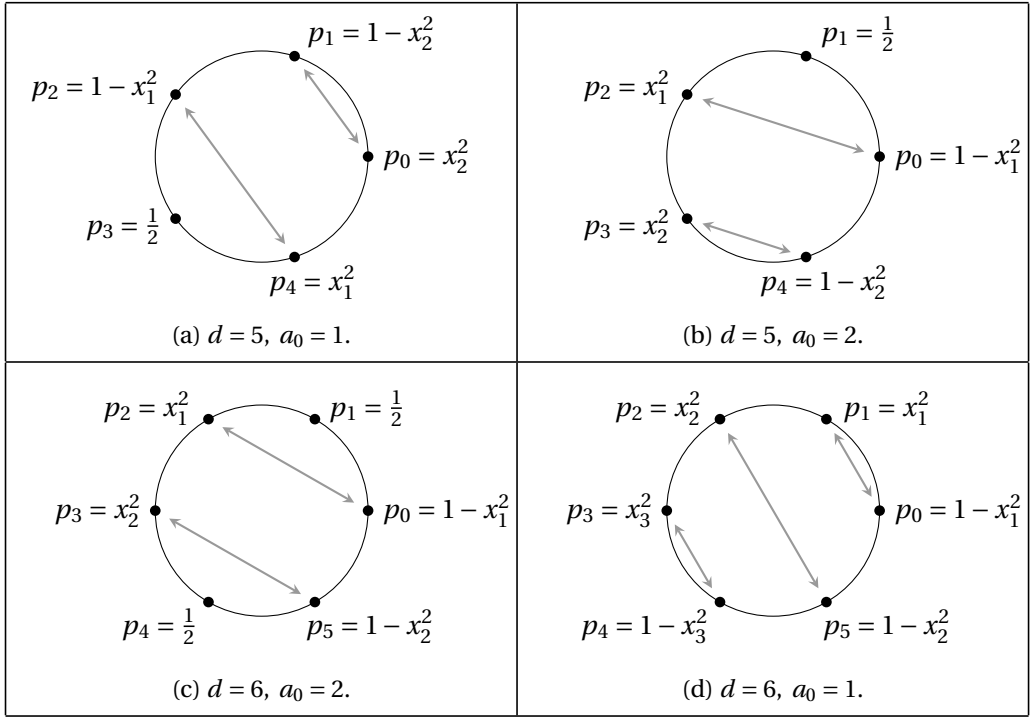
\begin{figure}[htbp]
  \centering
  \setlength{\arrayrulewidth}{0.6pt}

  \begin{tabular}{|@{}p{0.42\linewidth}@{}|@{}p{0.42\linewidth}@{}|}
    \hline
    \subcaptionbox{$d=5,~a_0=1$.\label{fig:a0oddodd}}[\linewidth]{%
      
      \begin{tikzpicture}[scale=1.4,>=stealth]
        \path[use as bounding box] (-2.35,-1.4) rectangle (2.35,1.45);
        \tikzset{markstyle/.style={draw=blue, line width=0.8pt, fill=none}}

        \draw (0,0) circle (1);

        \foreach \k in {0,...,4}{
          \coordinate (P\k) at ({cos(\k*72)},{sin(\k*72)});
          \fill (P\k) circle (0.05);
          \node (M\k) at ($ (P\k)!-1.2mm!(0,0) $) {};
        }

        \node[anchor=west,xshift=-3pt,yshift=0pt] at (M0) {$p_0=x_2^2$};
        \node[xshift=27pt,yshift=4pt] at (M1) {$p_1=1-x_2^2$};
        \node[anchor=east, xshift=1pt,yshift=0pt] at (M2) {$p_2=1-x_1^2$};
        \node[anchor=east,xshift=2pt,yshift=0pt] at (M3) {$p_3=\frac 12$};
        \node[xshift=5pt,yshift=-5pt] at (M4) {$p_4=x_1^2$};

        \doublearrow{P0}{P1}
        \doublearrow{P2}{P4}
      \end{tikzpicture}
    } &
    
    \subcaptionbox{$d=5,~a_0=2$.\label{fig:a0oddeven}}[\linewidth]{%
      
      \begin{tikzpicture}[scale=1.4,>=stealth]
        \path[use as bounding box] (-2.35,-1.4) rectangle (2.35,1.45);
        \tikzset{markstyle/.style={draw=blue, line width=0.8pt, fill=none}}

        \draw (0,0) circle (1);

        \foreach \k in {0,...,4}{
          \coordinate (P\k) at ({cos(\k*72)},{sin(\k*72)});
          \fill (P\k) circle (0.05);
          \node (M\k) at ($ (P\k)!-1.2mm!(0,0) $) {};
        }

        \node[anchor=west,xshift=-3pt,yshift=0pt] at (M0) {$p_0=1-x_1^2$};
        \node[xshift=18pt,yshift=4pt] at (M1) {$p_1=\frac 12$};
        \node[anchor=east, xshift=1pt,yshift=0pt] at (M2) {$p_2=x_1^2$};
        \node[anchor=east,xshift=1pt,yshift=0pt] at (M3) {$p_3=x_2^2$};
        \node[xshift=5pt,yshift=-5pt] at (M4) {$p_4=1-x_2^2$};

        \doublearrow{P0}{P2}
        \doublearrow{P3}{P4}
      \end{tikzpicture}
    } \\[1.8em]    
    \hline
    
    \subcaptionbox{$d=6,~a_0=2$.\label{fig:a0eveneven}}[\linewidth]{%
      
      \begin{tikzpicture}[scale=1.4,>=stealth]
        \path[use as bounding box] (-2.35,-1.4) rectangle (2.35,1.4);
        \tikzset{markstyle/.style={draw=blue, line width=0.8pt, fill=none}}

        \draw (0,0) circle(1);
        \foreach \k in {0,...,5}{
          \coordinate (P\k) at ({cos(\k*60)},{sin(\k*60)});
          \fill (P\k) circle (0.05);
          \node (M\k) at ($ (P\k)!-1.2mm!(0,0) $) {};
        }

        \node[anchor=west,xshift=-3pt,yshift=0pt] at (M0) {$p_0=1-x_1^2$};
        \node[xshift=17pt,yshift=0pt] at (M1) {$p_1=\frac 12$};
        \node[anchor=east, xshift=-1pt,yshift=0pt] at (M2) {$p_2=x_1^2$};
        \node[anchor=east,xshift=3pt,yshift=0pt] at (M3) {$p_3=x_2^2$};
        \node[xshift=-16pt,yshift=-5pt] at (M4) {$p_4=\frac 12$};
        \node[xshift=18pt,yshift=-5pt] at (M5) {$p_5=1-x_2^2$};

        \doublearrow{P0}{P2}
        \doublearrow{P3}{P5}
      \end{tikzpicture}
    } &
    
    \subcaptionbox{$d=6,~a_0=1$.\label{fig:a0evenodd}}[\linewidth]{%
      
      \begin{tikzpicture}[scale=1.4,>=stealth]
        \path[use as bounding box] (-2.35,-1.4) rectangle (2.35,1.4);
        \tikzset{markstyle/.style={draw=blue, line width=0.8pt, fill=none}}

        \draw (0,0) circle(1);
        \foreach \k in {0,...,5}{
          \coordinate (P\k) at ({cos(\k*60)},{sin(\k*60)});
          \fill (P\k) circle (0.05);
          \node (M\k) at ($ (P\k)!-1.2mm!(0,0) $) {};
        }

        \node[anchor=west,xshift=-3pt,yshift=0pt] at (M0) {$p_0=1-x_1^2$};
        \node[xshift=20pt,yshift=0pt] at (M1) {$p_1=x_1^2$};
        \node[anchor=east, xshift=-1pt,yshift=0pt] at (M2) {$p_2=x_2^2$};
        \node[anchor=east,xshift=3pt,yshift=0pt] at (M3) {$p_3=x_3^2$};
        \node[xshift=-26pt,yshift=-5pt] at (M4) {$p_4=1-x_3^2$};
        \node[xshift=20pt,yshift=-5pt] at (M5) {$p_5=1-x_2^2$};

        \doublearrow{P0}{P1}
        \doublearrow{P2}{P5}
        \doublearrow{P3}{P4}
      \end{tikzpicture}
    } \\[1.8em] 
    \hline
  \end{tabular}

   \caption{Illustration of the construction in Section~\ref{sec:xpl value of c} for several pairs $(d, a_0)$, with $\ZZd$ represented as $d$ points on a circle. The double arrows indicate points that are exchanged by the involution $\sigma$.}%
   \label{fig:a0constr}

\end{figure}

Fix an integer $d\geq3$, and identify $\cH=\CC^d$ with $\CC^\ZZd$, where $\ZZd=\ZZ/d\ZZ$ is  equipped with its 
additive group structure. Denote by $(e_a)_{a\in\ZZd}$ the canonical basis of $\cH$, and by 
$(e_a^\ast)_{a\in\ZZd}$ the dual basis. To the involution $\sigma:\ZZd\to\ZZd$,
given by $\sigma(a)=a_0-a$ for some $a_0\in\ZZd$, we associate the set
$$
\fP_{d,\sigma}=\left\{\bsp=(p_a)_{a\in\ZZd}\in[0,1]^\ZZd\mid p_a+p_{\sigma(a)}=1 \text{ for all }a\in\ZZd\right\}.
$$
There exists a smooth bijection 
$[0,1]^{\hat d}\ni\bsx=(x_k)_{k\in\llbracket1,\hat d\rrbracket}\mapsto\bsp=(p_a)_{a\in\ZZd}\in\fP_{d,\sigma}$,
where
\begin{itemize}
\item $\hat d=(d-1)/2$ if $d$ is odd, and
$$
p_{\hat a}=\tfrac12,\qquad p_{\hat a+k}=x_k^2=1-p_{\hat a-k}
$$
for all $k\in\llbracket1,\hat d\rrbracket$, with $\hat a=(a_0+d)/2$ if $a_0$ is odd  (see Figure~\ref{fig:a0oddodd}) and $\hat a=a_0/2$ if $a_0$ is even (see Figure~\ref{fig:a0oddeven});
\item $\hat d=(d-2)/2$ if $d$ and $a_0$ are even, and
$$
p_{\frac{a_0}2}=p_{\frac{a_0+d}2}=\tfrac12,\qquad p_{\frac{a_0}2+k}=x_k^2=1-p_{\frac{a_0}2-k}
$$
for all $k\in\llbracket1,\hat d\rrbracket$ (see Figure~\ref{fig:a0eveneven});
\item $\hat d=d/2$ if $d$ is even and $a_0$ is odd, and
$$
p_{\frac{a_0-1}2+k}=x_k^2=1-p_{\frac{a_0+1}2-k}
$$
for all $k\in\llbracket1,\hat d\rrbracket$ (see Figure~\ref{fig:a0evenodd}).
\end{itemize}

In the following, we equip $\fP_{d,\sigma}$ with the push-forward of the
Lebesgue measure on $[0,1]^{\hat d}$. We shall repeatedly use the fact that if
$F:{}]0,1[^{\hat d}\to\RR$ is a real analytic function
which does not vanish identically, then
the subset $\{\bsx\in{}]0,1[^{\hat d}\mid F(\bsx)\neq0\}$ has full measure 
(see~\cite[Proposition~1]{Mityagin2020}).

Given $\bseta=(\eta_a)_{a\in\ZZd}\in\TT^\ZZd$, we set
$$
J=K\left(\sum_{a\in\ZZd}\eta_a\,e_{\sigma(a-1)}\otimes e_a^\ast\right),
$$
where $K$ is either the identity or the complex conjugation in the canonical basis.
$J$ is thus either unitary or anti-unitary.

For $\bsp=(p_a)_{a\in\ZZd}\in\fP_{d,\sigma}$, we further set
$$
V_1=\sum_{a\in\ZZd}\sqrt{p_a}\,e_{a+1}\otimes e_a^\ast\qquad
V_2=j^{-1}(V_1^\ast)=\sum_{a\in\ZZd}\overline\eta_{a+1}\eta_a\sqrt{p_{\sigma(a)}}\,e_{a+1}\otimes e_a^\ast ,
$$
and define $\Phi:X\mapsto V_1^\ast X V_1+V_2^\ast X V_2$. For $a,b\in\ZZd$, one has
\beq
\Phi(e_a\otimes e_b^\ast)=C_{a,b}\,e_{a-1}\otimes e_{b-1}^\ast,\qquad
\Phi^\ast(e_a\otimes e_b^\ast)=C_{b+1,a+1}\,e_{a+1}\otimes e_{b+1}^\ast,
\label{eq:phimat}
\eeq
with 
$$
C_{a,b}=\sqrt{p_{a-1}p_{b-1}}
+(\eta_a\,\overline\eta_{a-1})(\overline \eta_b\,\eta_{b-1})
\sqrt{p_{\sigma(a-1)}p_{\sigma(b-1)}}
=\overline{C}_{b,a}.
$$
In particular $C_{a,a}=1$, so that
$$
V_1^\ast V_1+V_2^\ast V_2=V_1V_1^\ast+V_2V_2^\ast=\one,
$$
which shows that $\Phi$ is a quantum channel with faithful invariant state $\rho=\one/d$.

\bed
We denote by $\cC_{d,\sigma}$ the family of channels constructed in this way.
\eed

\bep\label{prop:C_one}
Consider the channels $\Phi$ in $\cC_{d,\sigma}$ as a function of $\bsp$, all the other parameters being arbitrary but fixed.
For almost all $\bsp\in\fP_{d,\sigma}$, $\Phi$  is irreducible. Moreover, any irreducible channel in $\cC_{d,\sigma}$  has period $d$, and maximal cycle $(P_{\xi_d^a})_{a\in\ZZd}$, given by $P_{\xi_d^a}=e_a\otimes e_a^\ast$.
\eep
\proof
Iterating~\eqref{eq:phimat}, we get
$$
\Phi^{nd}(e_a\otimes e_b^\ast)=\left(\prod_{r\in\ZZd}C_{a-r,b-r}\right)^n e_a\otimes e_b^\ast.
$$
The Cauchy--Schwarz inequality implies $|C_{a,b}|\leq\sqrt{p_{a-1}p_{b-1}}+\sqrt{p_{\sigma(a-1)}p_{\sigma(b-1)}}\le1$,
where the second inequality is  strict unless $p_{a-1}=p_{b-1}$. On the one hand, it follows that if $a\neq b$, then
$$
\left|\prod_{r\in\ZZd}C_{a-r,b-r}\right|<1,
$$
for almost all $(p_a)_{a\in\ZZd}\in\fP_{d,\sigma}$, and hence
$$
\lim_{n\to\infty}\Phi^n(e_a\otimes e_b^\ast)=0.
$$
Since on the other hand $\Phi(e_a\otimes e_a^\ast)=e_{a-1}\otimes e_{a-1}^\ast$, we conclude that
$$
\lim_{n\to\infty}\frac 1d\sum_{k=0}^{d-1}\Phi^{nd+k}(X)=\one\bra\rho,X\ket_\HS
$$
for all $X\in\cB(\cH)$. Invoking~\cite[Proposition~2.2]{Evans1987} we conclude that $\Phi$ is irreducible.
The last statements follow from the fact that $\Phi(P_{\xi_d^a})=P_{\xi_d^{a-1}}$ for any $a\in\ZZd$.\QED

\bep\label{prop:C_two}
Let $\Phi\in \cC_{d,\sigma}$ be an irreducible channel. Then $\Phi$ satisfies~\QDB{} with $\Phi(J^2)=\eta J^2$ iff
$\eta\in\TT_d$ and, for some $\zeta\in\TT$ and all $a\in\ZZd$, 
$$
\eta^\#_{\sigma(a-1)}\eta_a=\zeta\eta^a,
$$
where for $z\in\CC$, we set
$z^\#=J^\ast zJ\in\CC$, \ie $z^\#=z$ if $J$ is unitary and $z^\#=\overline z$ if $J$ is anti-unitary.
\eep
\proof Since $\rho$ is a multiple of the identity, one has $j(\rho)=\rho$ and $\Phi^\rho=\Phi^\ast$. Thus,
setting $\theta_a=\eta_{\sigma(a-1)}^\#\eta_a$, we have to show that $\theta_a=\zeta\eta^a$
for all $a\in\ZZd$ is a necessary and sufficient condition for both relations
\beq
\Phi(J^2)=\eta J^2,\qquad
j^{-1}\circ\Phi^\ast\circ j=\Phi,
\label{eq:tworel}
\eeq
to hold.

Concerning the first relation, we observe that by Proposition~\ref{prop:C_one}, $\Phi$ has period $d$, and Proposition~\ref{prop:prop channel} gives $\eta\in\TT_d$. A simple calculation yields
$$
J^2=\sum_{a\in\ZZd}\theta_a\,e_a\otimes e_a^\ast,\qquad
\Phi(J^2)=\sum_{a\in\ZZd}\theta_{a+1}\,e_a\otimes e_a^\ast,
$$
so that $\Phi(J^2)=\eta J^2$ holds iff $\theta_{a+1}=\eta\theta_a$ for all $a\in\ZZd$,
and hence $\theta_a=\zeta\eta^a$ for some $\zeta\in\TT$.

It remains to show that the last relation implies the second one in~\eqref{eq:tworel}.
Since
$$
j^{-1}\circ\Phi^\ast\circ j(X)=j^{-1}(V_1)Xj^{-1}(V_1^\ast)+j^{-1}(V_2)Xj^{-1}(V_2^\ast),
$$
and, by definition of $\cC_{d,\sigma}$, $j^{-1}(V_1^\ast)=V_2$, it suffices to show that
$j^{-1}(V_2^\ast)=\lambda V_1$ for some $\lambda\in\TT$, or equivalently $j^{-2}(V_1)=\lambda V_1$.
Another simple calculation gives
$$
J^{2\ast}V_1J^2=\sum_{a\in\ZZd}\overline{\theta}_{a+1}\theta_a\sqrt{p_a}\,e_{a+1}\otimes e_a^\ast
=\overline{\eta}V_1,
$$
and ends the proof.\QED

Let $\Gamma$ be a unitary or anti-unitary operator on $\cH$. A channel $\Phi\in\CPU(\cH)$ is said to be
$\Gamma$-covariant if, for all $X\in\cB(\cH)$, $\Phi(\Gamma X\Gamma^\ast)=\Gamma\Phi(X)\Gamma^\ast$.

\begin{proposition}\label{prop:general example}
Consider the channels $\Phi$ in $\cC_{d,\sigma}$ 
as a function of $\bsp$, all the other parameters being arbitrary but fixed.
Assume that for some $\bsp\in\fP_{d,\sigma}$ and any $r\in\ZZd\setminus\{0\}$,
$|C_{a-r,b-r}|\neq|C_{a,b}|$ for some $a,b\in\ZZd$. Then, the following hold for almost all 
$\bsp\in\fP_{d,\sigma}$:
\begin{enumerate}[label=(\roman*)]
\item If\, $\Phi$ is $\Gamma$-covariant for some unitary $\Gamma$, 
then $\Gamma$ is a peripheral eigenvector of\, $\Phi$, \ie
$$
\Gamma=z\sum_{a\in\ZZd}\xi_d^{na}e_a\otimes e_a^\ast
$$
for some $z\in\TT$ and $n\in\ZZ$. 
\item If\, $\Phi$ is $\Gamma$-covariant for some anti-unitary $\Gamma$, then 
$$
\Gamma e_a=z_ae_a
$$
for some $(z_a)_{a\in\ZZd}\in\TT^{\ZZd}$, and in particular $\Gamma^2=\one$. Moreover, for any integer $n\geq1$ and any $b_0,\ldots,b_n\in\ZZd$ such that $b_n=b_0$,
\begin{equation}\label{eq:productwalk}
\prod_{k=0}^{n-1}C_{b_k,b_{k+1}}\in\RR.
\end{equation}
\end{enumerate}
\end{proposition}

\proof
For any $r\in\ZZd\setminus\{0\}$, consider the real analytic function
$$
{}]0,1[^{\hat d}\ni \bsx
\mapsto F_r=\sum_ {a,b\in\ZZd}
\left(|C_{a-r,b-r}|^2-|C_{a,b}|^2\right)^2.
$$
Since $F_r$ extends continuously to $[0,1]^{\hat d}$, the hypothesis implies that none of the functions $F_r$, $r\neq0$, vanishes identically.
By real analyticity and finiteness, there exists a full-measure set
$\Omega\subset\fP_{d,\sigma}$ such that, for every $\bsp\in\Omega$,
$$
F_r(\bsp)>0\quad(r\neq0).
$$
Shrinking $\Omega$ by another null set, we may also assume that $\Phi$
satisfies the conclusions of Proposition~\ref{prop:C_one}. We fix an
arbitrary $\bsp\in\Omega$.

\medskip\noindent
{\sl (i)}
The $\Gamma$-covariance of $\Phi$ yields that, for any $a\in\ZZd$,
$$
\Phi(\Gamma P_{\xi_d^a}\Gamma^\ast)
=\Gamma\Phi(P_{\xi_d^a})\Gamma^\ast
=\Gamma P_{\xi_d^{a-1}}\Gamma^\ast.
$$
Since $(\Gamma P_{\xi_d^a}\Gamma^\ast)_{a\in\ZZd}$ is a partition of unity,
the uniqueness of the maximal cycle of $\Phi$ implies that
$\Gamma P_{\xi_d^a}\Gamma^\ast=P_{\xi_d^{a-r}}$ for some $r\in\ZZd$.
Using the fact that
$\Gamma P_{\xi_d^a}\Gamma^\ast=(\Gamma e_a)\otimes(\Gamma e_a)^\ast$,
we derive $\Gamma e_a=z_ae_{a-r}$ for some
$(z_a)_{a\in\ZZd}\in\TT^{\ZZd}$. Thus, for any $a,b\in\ZZd$,
\begin{align*}
  \Phi(\Gamma e_a\otimes e_b^\ast\Gamma^\ast)
&=C_{a-r,b-r}z_a\overline z_b\,e_{a-r-1}\otimes e_{b-r-1}^\ast\\
&=C_{a,b}z_{a-1}\overline z_{b-1}\,e_{a-r-1}\otimes e_{b-r-1}^\ast\\
&=\Gamma\Phi(e_a\otimes e_b^\ast)\Gamma^\ast.
\end{align*}
It follows that $|C_{a-r,b-r}|=|C_{a,b}|$ for all $a,b\in\ZZd$.
Since $\bsp\in\Omega$, this forces $r=0$.

Set $u_a=z_a\overline z_{a-1}$. Then, for any $a,b\in \ZZd$, 
$$C_{a,b}u_a=C_{a,b}u_b.$$
Hence, if for any $a,b\in \ZZd$, there exists $c\in \ZZd$ such that $C_{a,c}C_{c,b}\neq 0$, then there exists $\zeta\in \CC$ of modulus $1$ such that $u_a=\zeta$ for any $a\in \ZZd$. Therefore
$z_a=z_0\zeta^a$, and
$$
\Gamma=z_0\sum_{a\in\ZZd}\zeta^ae_a\otimes e_a^\ast.
$$
It follows that $\Phi(\Gamma)=\zeta\Gamma$, which, by
Proposition~\ref{prop:prop channel}, implies that
$\zeta=\xi_d^k$ for some $k\in\ZZd$.

It remains to prove that for almost all $\bsp\in \fP_{d,\sigma}$, for all $a,b\in \ZZd$ there exists $c\in \ZZd$ such that $C_{a,c}C_{c,b}\neq 0$. Since the function $$]0,1[^{\hat d}\ni \bsx
\mapsto \widetilde{F}=\prod_{a,b\in \ZZd} \sum_{c\in \ZZd}|C_{a,c}C_{c,b}|^2$$
is real analytic, it is sufficient to prove it does not vanish identically.

The map $\sigma$ has at most two fixed points, and since $d\geq 3$ by assumption, there exists some $a^*\in\ZZd$ such that $\sigma(a^*)\neq a^*$. Let $\bsq\in \fP_{d,\sigma}$ be given by $q_{a^*}=\frac14$, $q_{\sigma(a^*)}=\frac34$, and $q_a=\frac12$ for all $a\notin \{a^*, \sigma(a^*)\}$. Set $A=a^*+1$ and $B=\sigma(a^*)+1$. For any $a\in\ZZd\setminus\{A,B\}$, we have $C_{a,A}\neq 0$, since
$$\sqrt{q_{a^*}q_{a-1}}=\frac{1}{2\sqrt{2}}, \quad 
\sqrt{(1-q_{a^*})(1-q_{a-1})}=\frac{\sqrt{3}}{2\sqrt{2}},$$ and similarly
$C_{a,B}\neq 0$.  Since $C_{A,A}=C_{B,B}=1$ and $C_{a,b}=\overline{C}_{b,a}$, given
$a,b\in\ZZd$, we indeed obtain $C_{a,c}C_{c,b}\neq0$ with
$$
c=\begin{cases}
A,& \text{if } B\notin\{a,b\},\\
B,& \text{if } B\in\{a,b\}\text{ and }A\notin\{a,b\},\\
c_0,& \text{if } \{a,b\}=\{A,B\},
\end{cases}
$$
where  $c_0\in\ZZd\setminus\{A,B\}$ is arbitrary.
Thus, as claimed, $\widetilde F$ does not vanish identically.

\medskip\noindent
{\sl (ii)} Proceeding as before, we have again $\Gamma e_a=z_a e_{a-r}$ for some $r\in\ZZd$,
$(z_a)_{a\in\ZZd}\in\TT^\ZZd$ and all $a\in\ZZd$. It follows that
\begin{equation}\label{eq:phiGantilin}
\begin{split}
\Phi(\Gamma e_a\otimes e_b^\ast\Gamma^\ast)&=C_{a-r,b-r}z_a\overline{z}_b\,e_{a-r-1}\otimes e_{b-r-1}^\ast\\
&=\overline{C}_{a,b}z_{a-1}\overline{z}_{b-1}\,e_{a-r-1}\otimes e_{b-r-1}^\ast\\
&=\Gamma\Phi(e_a\otimes e_b^\ast)\Gamma^\ast,
\end{split}
\end{equation}
which again leads to $r=0$ for almost all $\bsp\in\fP_{d,\sigma}$. Hence, for such $\bsp$,
we have 
$$
\Gamma^2e_a=\Gamma z_a e_a=\overline{z}_a\Gamma e_a=\overline{z}_az_a e_a=e_a
$$
and thus $\Gamma^2=\one$.

Setting $r=0$ in \eqref{eq:phiGantilin} and identifying the coefficients, we get
$$
C_{a,b}z_a\overline z_b
=\overline C_{a,b}\,z_{a-1}\overline z_{b-1}.
$$
Let $n\geq1$ and let $\alpha_0,\ldots,\alpha_n\in\ZZd$ be such that
$\alpha_n=\alpha_0$. Applying this identity with $(a,b)=(\alpha_k,\alpha_{k+1})$ and multiplying over $k\in\llbracket0,n-1\rrbracket$, the phases cancel on both sides, and we obtain
$$
\prod_{k=0}^{n-1}C_{\alpha_k,\alpha_{k+1}}
=
\prod_{k=0}^{n-1}\overline C_{\alpha_k,\alpha_{k+1}}.
$$
Therefore, \eqref{eq:productwalk} holds.
\QED

\medskip
We now prove Propositions~\ref{prop: no unitary no 1}--\ref{prop:unitary no 1} using channels in $\cC_{d,\sigma}$.

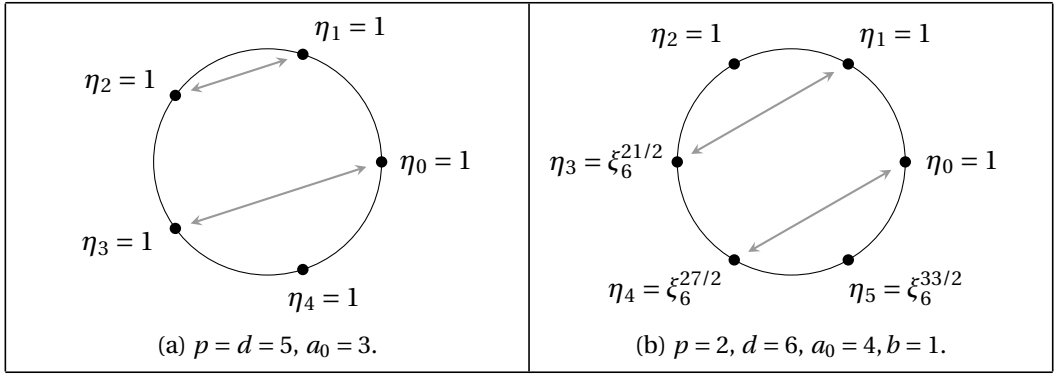
\begin{figure}[htbp]
  \centering
  \setlength{\arrayrulewidth}{0.6pt}

  \begin{tabular}{|@{}p{0.43\linewidth}@{}|@{}p{0.43\linewidth}@{}|}
    \hline

    \subcaptionbox{$p=d=5$, $a_0=3$.\label{fig:dp5_a0_3_propI}}[\linewidth]{%
      \begin{tikzpicture}[scale=1.5,>=stealth]
		\path[use as bounding box] (-2.15,-1.35) rectangle (2.15,1.4);
        \tikzset{markstyle/.style={draw=blue, line width=0.8pt, fill=none}}
        \draw (0,0) circle (1);

        \foreach \k in {0,...,4}{
          \coordinate (P\k) at ({cos(\k*72)},{sin(\k*72)});
          \fill (P\k) circle (0.05);
          \node (M\k) at ($ (0,0)!1.22!(P\k) $) {};
        }

        \node[anchor=west,xshift=-6pt,yshift=0pt]  at (M0) {$\begin{aligned}\eta_0&=1\end{aligned}$};
        \node[xshift=20pt,yshift=1pt]              at (M1) {$\begin{aligned}\eta_1&=1\end{aligned}$};
        \node[anchor=east, xshift=14pt,yshift=0pt] at (M2) {$\begin{aligned}\eta_2&=1\end{aligned}$};
        \node[anchor=east, xshift=13pt,yshift=0pt] at (M3) {$\begin{aligned}\eta_3&=1\end{aligned}$};
        \node[xshift=10pt,yshift=-2pt]             at (M4) {$\begin{aligned}\eta_4&=1\end{aligned}$};

        \doublearrow{P0}{P3}
        \doublearrow{P1}{P2}
      \end{tikzpicture}
    } &

    \subcaptionbox{$p=2$, $d=6$, $a_0=4, b=1$.\label{fig:d6_a0_4_propI}}[\linewidth]{%
      \begin{tikzpicture}[scale=1.5,>=stealth]
		\path[use as bounding box] (-2.15,-1.35) rectangle (2.15,1.4);
        \tikzset{markstyle/.style={draw=blue, line width=0.8pt, fill=none}}
        \draw (0,0) circle (1);

        \foreach \k in {0,...,5}{
          \coordinate (P\k) at ({cos(\k*60)},{sin(\k*60)});
          \fill (P\k) circle (0.05);
          \node (M\k) at ($ (0,0)!1.22!(P\k) $) {};
        }

        \node[anchor=west,xshift=-5pt,yshift=0pt]  at (M0) {$\begin{aligned}\eta_0&=1\end{aligned}$};
        \node[xshift=18pt,yshift=3pt]             at (M1) {$\begin{aligned}\eta_1&=1\end{aligned}$};
        \node[xshift=-8pt,yshift=3pt]             at (M2)  {$\begin{aligned}\eta_2&=1\end{aligned}$};
        \node[anchor=east, xshift=18pt,yshift=0pt]  at (M3) {$\begin{aligned}\eta_3&=\xi_6^{21/2}\end{aligned}$};
        \node[xshift=-16pt,yshift=-3pt]           at (M4) {$\begin{aligned}\eta_4&=\xi_6^{27/2}\end{aligned}$};
        \node[xshift=22pt,yshift=-3pt]            at (M5) {$\begin{aligned}\eta_5&=\xi_6^{33/2}\end{aligned}$};

        \doublearrow{P0}{P4}
        \doublearrow{P1}{P3}
      \end{tikzpicture}
    } \\[1.8em]

    \hline
  \end{tabular}

  \caption{Illustration of the proof of Proposition~\ref{prop: no unitary no 1}(i).}
  \label{fig:propI_side_by_side}
\end{figure}

\noindent{\bf Proof of Proposition~\ref{prop: no unitary no 1}}.
\textsl{(i)} Let $\eta\in\TT_p$, $p\ge2$. Assuming first that $p$ is odd,
consider irreducible channels $\Phi$ in $\cC_{d,\sigma}$ with $d=p$,
$\eta_a=1$ for $a\in\ZZ_d$, and arbitrary $a_0\in\ZZ_d$; see Figure~\ref{fig:dp5_a0_3_propI}. Then, by Proposition~\ref{prop:C_two}, 
$(\Phi,\rho)$ satisfies \QDB{} with respect to $J$,
and $\Phi(J^2)=J^2$. The claim now follows from 
Propositions~\ref{prop:C_one} and~\ref{prop:possible values of c}(ii).

To deal with an even $p$, let $\eta=\xi_p^b$ for some $b\in\ZZ_p$, and
consider any irreducible channel $\Phi$ in $\cC_{d,\sigma}$ with $d=3p\geq6$,
$a_0=d-2$, $\hat d=(d-2)/2$ and
$$
\eta_a=\begin{cases}
1&\text{for }a\in\llbracket0,\hat d\rrbracket,\\[4pt]
\xi_d^{3b(a+1/2)}&\text{for }a\in\llbracket\hat d+1,d-1\rrbracket,
\end{cases} 
$$
where $\xi_d^t:=\e^{2\pi\i t/d}$ for $t\in\RR$, as depicted in Figure~\ref{fig:d6_a0_4_propI}.
Direct computation leads to
$$
\overline{\eta}_{\sigma(a-1)}\eta_a=\xi_d^{3b/2}\xi_d^{3ba}
$$
for all $a\in\ZZd$, and Propositions~\ref{prop:C_one}~and~\ref{prop:C_two} imply
that almost all $\Phi\in \cC_{d,\sigma}$ satisfy \QDB{} with $J$ anti-unitary such that 
$\Phi(J^2)=\xi_d^{3b}J^2=\eta J^2$.

\begin{figure}[htbp]
\centering
\begin{tikzpicture}[scale=2.7,>=stealth]
\tikzset{markstyle/.style={draw=blue, line width=0.8pt, fill=none}} 

\draw[] (0,0) circle(1);

\foreach \k in {0,...,11}{
  \coordinate (P\k) at ({cos(\k*30)},{sin(\k*30)});
  \fill (P\k) circle (0.035);
  \node (M\k) at ($ (P\k)!-1.2mm!(0,0) $) {};
}
  \node[anchor=west,xshift=-4pt,yshift=0pt] at (M0) {$\begin{aligned}p_0&=s\\[-0.4em]\eta_0&=1\end{aligned}$};
  \node[xshift=15pt,yshift=1pt] at (M1) {$\begin{aligned}p_1&=1\\[-0.4em]\eta_1&=1\end{aligned}$};
  \node[anchor=east, xshift=25pt,yshift=11pt] at (M2) {$\begin{aligned}p_2&=1\\[-0.4em]\eta_2&=1\end{aligned}$};
  \node[anchor=east,xshift=15pt,yshift=9pt] at (M3) {$\begin{aligned}p_3&=1\\[-0.4em]\eta_3&=1\end{aligned}$};
  \node[xshift=-13pt,yshift=6pt] at (M4) {$\begin{aligned}p_4&=1\\[-0.4em]\eta_4&=1\end{aligned}$};
  \node[xshift=-14pt,yshift=1pt] at (M5) {$\begin{aligned}p_5&=\tfrac 12\\[-0.4em]\eta_5&=1\end{aligned}$};
  \node[xshift=-18pt,yshift=1pt] at (M6) {$\begin{aligned}p_6&=0\\[-0.4em]\eta_6&=\xi_{12}^{39/2}\end{aligned}$};
  \node[xshift=-17pt,yshift=-1pt] at (M7) {$\begin{aligned}p_7&=0\\[-0.4em]\eta_7&=\xi_{12}^{45/2}\end{aligned}$};
  \node[xshift=-11pt,yshift=-5pt] at (M8) {$\begin{aligned}p_8&=0\\[-0.4em]\eta_8&=\xi_{12}^{51/2}\end{aligned}$};
  \node[xshift=6pt,yshift=-8pt] at (M9) {$\begin{aligned}p_9&=0\\[-0.4em]\eta_9&=\xi_{12}^{57/2}\end{aligned}$};
  \node[xshift=20pt,yshift=-9pt] at (M10) {$\begin{aligned}p_{10}&=1-s\\[-0.4em]\eta_{10}&=\xi_{12}^{63/2}\end{aligned}$};
  \node[xshift=23pt,yshift=-3pt] at (M11) {$\begin{aligned}p_{11}&=\tfrac 12\\[-0.4em]\eta_{11}&=\xi_{12}^{69/2}\end{aligned}$};

\doublearrow{P0}{P10}
\doublearrow{P1}{P9}
\doublearrow{P2}{P8}
\doublearrow{P3}{P7}
\doublearrow{P4}{P6}

\end{tikzpicture}

  \caption{Illustration of the values of $p_a$ and $\eta_a$ in Table~\ref{tab:valuespaeta} with $p=4, d=12,~a_0=10$.}
  \label{fig:d12constr}
\end{figure}
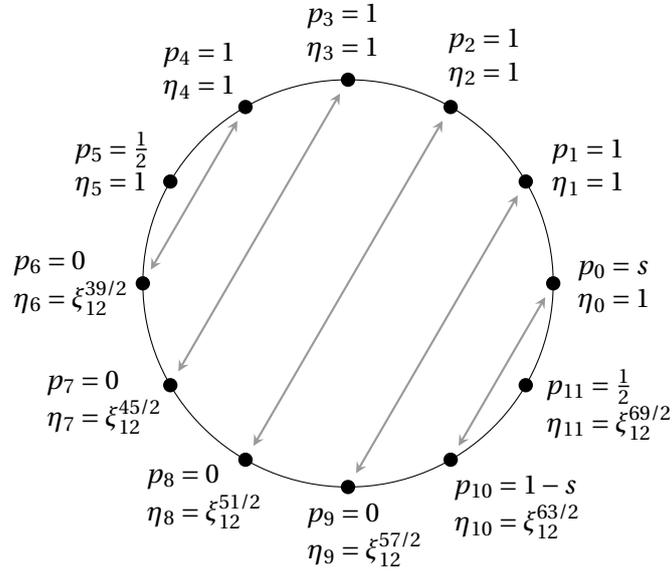

\noindent\textsl{(ii)} Fix an even $p\geq4$ and consider the previous set of channels $\cC_{d,\sigma}$ with $b=1$. We first show that the assumption
of Proposition~\ref{prop:general example} holds. For this, consider $\bsp$
as given in Table~\ref{tab:valuespaeta}, with $s\in (0, 1/2)$. The case $p=4$ is illustrated in Figure~\ref{fig:d12constr}.

\begin{table}[htbp]
\begin{center}
\renewcommand{\arraystretch}{1.2}
\begin{tabular}{|c|c|c|c|c|c|c|}
	\hline
	$a$ & $\eta_a$ & $p_a$ & $p_{a - 1}\cdot p_a$ & $\overline{\eta}_{a - 1}
	\eta_a^2 \overline{\eta}_{a + 1}$ & $p_{\sigma (a - 1)} \cdot p_{\sigma (a)}$&$C_{a,a+1}$\\
	\hline
	$0$ & $1$ & $s$ & $\tfrac{1}{2} \cdot s$ & $\xi_d^{3 / 2}$ & $\tfrac{1}{2}
	\cdot (1 - s)$&$\sqrt{s/2}+\xi_d^{3/2}\sqrt{(1-s)/2}$\\
	$1$ & $1$ & $1$ & $s \cdot 1$ & $1$ & $(1 - s) \cdot 0$&$\sqrt{s}$\\
	$2$ & $1$ & $1$ & $1 \cdot 1$ & $1$ & $0 \cdot 0$&$1$\\
	{\vdots} & {\vdots} & {\vdots} & {\vdots} & {\vdots} & {\vdots}& {\vdots}\\
	$\hat{d} - 1$ & $1$ & $1$ & $1 \cdot 1$ & $1$ & $0 \cdot 0$&$1$\\
	$\hat{d}$ & $1$ & $1 / 2$ & $1 \cdot \tfrac{1}{2}$ & $- \xi_d^{ -3 / 2}$ &
	$\tfrac{1}{2} \cdot 0$&$\sqrt{1/2}$\\
	$\hat{d} + 1$ & $\xi_d^{3 (\hat{d} + 3 / 2)}$ & $0$ & $
	\tfrac{1}{2}\cdot 0$ & $- \xi_d^{- 3 / 2}$ & $1 \cdot \tfrac{1}{2}$&$-\xi_d^{-3/2}\sqrt{1/2}$\\
	$\hat{d} + 2$ & $\xi_d^{3 (\hat{d} + 5 / 2)}$ & $0$ & $0 \cdot 0$ & $1$ &
	$1 \cdot 1$&$1$\\
	{\vdots} & {\vdots} & {\vdots} & {\vdots} & {\vdots} & {\vdots}& {\vdots}\\
	$d - 3$ & $\xi_d^{3 (d - 5 / 2)}$ & $0$ & $0 \cdot 0$ & $1$ & $1 \cdot
	1$&$1$\\
	$d - 2$ & $\xi_d^{3 (d - 3 / 2)}$ & $1 - s$ & $0 \cdot (1 - s)$ & $1$ & $1
	\cdot s$&$\sqrt{s}$\\
	$d - 1$ & $\xi_d^{3 (d - 1 / 2)}$ & 1/2 & $(1 - s) \cdot \tfrac{1}{2}$ &
	$\xi_d^{3 / 2}$ & $s \cdot \tfrac{1}{2}$&$\sqrt{(1-s)/2}+\xi_d^{3/2}\sqrt{s/2}$\\
	\hline
\end{tabular}
\end{center}\caption{Values of $\eta_a$, $p_a$, and derived quantities in the proof of Proposition~\ref{prop: no unitary no 1}(ii).}\label{tab:valuespaeta}
\end{table}

Observe that
\begin{itemize}
\item for $r=1$, 
$$
|C_{\hat d-1,\hat d}|=1>\sqrt{1/2}=|C_{\hat d,\hat d+1}|=|C_{\hat d-1+r,\hat d+r}|;
$$
\item for $r\in\llbracket2,\hat d-1\rrbracket\cup\llbracket\hat d+2,d-3\rrbracket$,
$$
|C_{0,1}|=\left|\sqrt{s/2}+\xi_d^{3/2}\sqrt{(1-s)/2}\right|<\cos\tfrac{\pi}{2p}<1=|C_{0+r,1+r}|;
$$
\item for $r\in\{\hat d,\hat d+1\}$,
$$
|C_{1,2}|=\sqrt{s}<\sqrt{1/2}=|C_{1+\hat d,2+\hat d}|
<1=|C_{1+\hat d+1,2+\hat d+1}|;
$$
\item for $r\in\{d-2,d-1\}$,
$$
|C_{2,3}|=1>\cos\tfrac\pi{2p}>|C_{0,1}|=|C_{2+d-2,3+d-2}|,
$$
and
$$
|C_{2,3}|=1>\sqrt{s}=|C_{1,2}|=|C_{2+d-1,3+d-1}|.
$$
\end{itemize}
Thus, $\cC_{d,\sigma}$ indeed satisfies the assumption of Proposition~\ref{prop:general example}.

Hence, the conclusions of  Propositions~\ref{prop:C_one}~and~\ref{prop:general example} hold for almost all $\bsp\in \fP_{d,\sigma}$, and we now fix such a $\bsp$. By Proposition~\ref{prop:C_two}, we then have $\Phi(J^2)=\xi_pJ^2$.

Suppose that there exists an anti-unitary operator $G$ such that $\Phi(G^2)=G^2$ and
$$
G^\ast\Phi^\rho(GXG^\ast)G=\Phi(X)
$$
for any $X\in\cB(\cH)$. Since $\Phi$ is irreducible, $G^2$ is proportional to the identity. The \QDB{} condition for $J$ and $G$ implies that
$\Phi$ is covariant w.r.t.\;$\Gamma=J^\ast G$. Proposition~\ref{prop:general example}(i) yields
that the unitary $\Gamma$ satisfies $\Gamma e_a=z\xi_d^{na}e_a$
for some $n\in\ZZ$ and $z\in\TT$. Since $J=KU$ with
$Ue_a=\eta_ae_{\sigma(a-1)}$, one has
$$
Ge_a=J\Gamma e_a=\overline z\,\xi_d^{-na}\overline\eta_a e_{\sigma(a-1)},
$$
and therefore
$$
G^2e_a=\xi_d^{n(a-\sigma(a-1))}\eta_a\overline\eta_{\sigma(a-1)}e_a
=\xi_d^{n+3/2}\xi_d^{(2n+3)a}e_a.
$$
Since $G^2$ is proportional to the identity, $d$ must divide $2n+3$. This is impossible because $d=3p$ is even whereas $2n+3$ is odd. This contradiction concludes the proof.
\QED

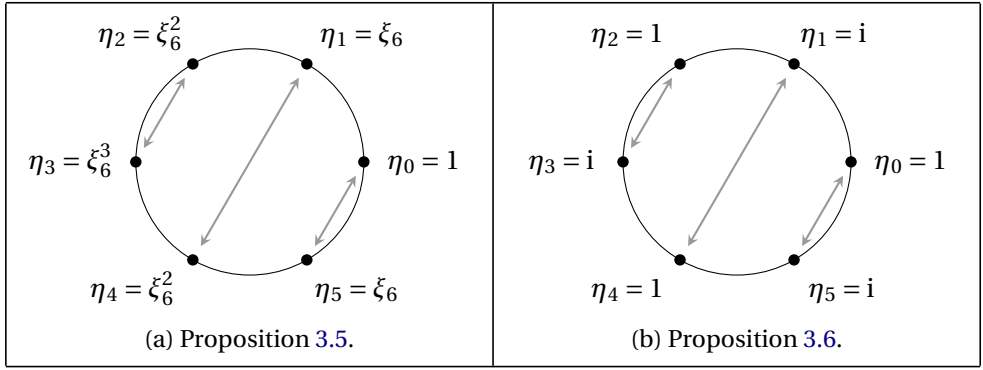
\begin{figure}[htb]
  \centering
  \setlength{\arrayrulewidth}{0.6pt}

  \begin{tabular}{|@{}p{0.4\linewidth}@{}|@{}p{0.4\linewidth}@{}|}
    \hline

    \subcaptionbox{Proposition~\ref{prop:anti U identity nu unitary}.\label{fig:d6_antiU_identity}}[\linewidth]{%
      \begin{tikzpicture}[scale=1.5,>=stealth]
        \path[use as bounding box] (-2.0,-1.3) rectangle (2.1,1.4);
        \tikzset{markstyle/.style={draw=blue, line width=0.8pt, fill=none}}
        \draw (0,0) circle (1);

        \foreach \k in {0,...,5}{
          \coordinate (P\k) at ({cos(\k*60)},{sin(\k*60)});
          \fill (P\k) circle (0.05);
          \node (M\k) at ($ (0,0)!1.22!(P\k) $) {};
        }

        \node[anchor=west, xshift=-4pt,yshift=0pt] at (M0) {$\begin{aligned}\eta_0&=1\end{aligned}$};
        \node[xshift=21pt,yshift=3pt] at (M1) {$\begin{aligned}\eta_1&=\xi_6\end{aligned}$};
        \node[xshift=-10pt,yshift=3pt] at (M2) {$\begin{aligned}\eta_2&=\xi_6^{2}\end{aligned}$};
        \node[anchor=east, xshift=14pt,yshift=0pt] at (M3) {$\begin{aligned}\eta_3&=\xi_6^{3}\end{aligned}$};
        \node[xshift=-13pt,yshift=-3pt] at (M4) {$\begin{aligned}\eta_4&=\xi_6^{2}\end{aligned}$};
        \node[xshift=19pt,yshift=-3pt] at (M5) {$\begin{aligned}\eta_5&=\xi_6\end{aligned}$};

        \doublearrow{P0}{P5}
        \doublearrow{P1}{P4}
        \doublearrow{P2}{P3}
      \end{tikzpicture}
    } &

    \subcaptionbox{Proposition~\ref{prop:unitary no 1}.\label{fig:d6_unitary_no1}}[\linewidth]{%
      \begin{tikzpicture}[scale=1.5,>=stealth]
        \path[use as bounding box] (-2.0,-1.3) rectangle (2.0,1.4);
        \tikzset{markstyle/.style={draw=blue, line width=0.8pt, fill=none}}
        \draw (0,0) circle (1);

        \foreach \k in {0,...,5}{
          \coordinate (P\k) at ({cos(\k*60)},{sin(\k*60)});
          \fill (P\k) circle (0.05);
          \node (M\k) at ($ (0,0)!1.22!(P\k) $) {};
        }

        \node[anchor=west, xshift=-6pt,yshift=0pt] at (M0) {$\begin{aligned}\eta_0&=1\end{aligned}$};
        \node[xshift=16pt,yshift=2pt] at (M1) {$\begin{aligned}\eta_1&=\i\end{aligned}$};
        \node[xshift=-8pt,yshift=2pt] at (M2) {$\begin{aligned}\eta_2&=1\end{aligned}$};
        \node[anchor=east, xshift=14pt,yshift=0pt] at (M3) {$\begin{aligned}\eta_3&=\i\end{aligned}$};
        \node[xshift=-8pt,yshift=-2pt] at (M4) {$\begin{aligned}\eta_4&=1\end{aligned}$};
        \node[xshift=16pt,yshift=-2pt] at (M5) {$\begin{aligned}\eta_5&=\i\end{aligned}$};

        \doublearrow{P0}{P5}
        \doublearrow{P1}{P4}
        \doublearrow{P2}{P3}
      \end{tikzpicture}
    } \\[1.8em]
    \hline
  \end{tabular}

  \caption{Illustration of the proofs of Propositions~\ref{prop:anti U identity nu unitary} and~\ref{prop:unitary no 1} with $d=6$, $a_0=5$.}
  \label{fig:d6_triple_array}
\end{figure}

{\bf Proof of Proposition~\ref{prop:anti U identity nu unitary}}.
Consider the set $\cC_{d,\sigma}$ with even $d\geq 6$, $a_0=d-1$, $\hat d=d/2$, and anti-unitary $J$. 
Set
$$
\eta_a=\begin{cases}
\xi_d^{a}&\text{for }a\in\llbracket0,\hat d-1\rrbracket;\\[4pt]
\xi_d^{\sigma(a-1)}&\text{for }a\in\llbracket\hat d,d-1\rrbracket;
\end{cases}
$$ 
so that $\overline{\eta}_{\sigma(a-1)}\eta_a=1$ for all $a\in\ZZd$; see Figure~\ref{fig:d6_antiU_identity}.

Direct computations lead to 
$$
C_{a,a+1}=\sqrt{p_a(1-p_{d-a})}+\sqrt{(1-p_a)p_{d-a}}
$$ 
for $a\in\ZZd\setminus\{0,\hat d\}$ and 
$$
C_{0,1}=2\xi_d^{-1}\cos\left(\frac{2\pi}{d}\right)\sqrt{p_0(1-p_0)},\qquad
C_{\hat d,\hat d+1}=2\xi_d\cos\left(\frac{2\pi}{d}\right)\sqrt{p_{\hat d}(1-p_{\hat d})}.
$$
The real analytic functions $]0,1[^{\hat d}\ni\bsx\mapsto (|C_{r,r+1}|^2-|C_{0,1}|^2)^2$
 clearly do not vanish identically for $r\neq0$. At the point $\bsp=\left(\frac 14,\frac 12,\ldots,\frac 12,\frac 34\right)\in\fP_{d,\sigma}$, we obtain
$$
\Im\left(C_{0,1}C_{1,2}C_{2,0}\right)
=-\frac{\sqrt3}{8}\sin\left(\frac{2\pi}{d}\right)\cos\left(\frac{2\pi}{d}\right)\neq0.
$$
Thus, the real analytic function $]0,1[^{\hat d}\ni\bsx\mapsto\Im(C_{0,1}C_{1,2}C_{2,0})$ does not vanish identically. Hence, there exists a full-measure set $\Omega\subset\fP_{d,\sigma}$ such that, for every $\bsp\in\Omega$, one has $|C_{r,r+1}|\neq|C_{0,1}|$ for all $r\in\ZZd\setminus\{0\}$ and $C_{0,1}C_{1,2}C_{2,0}\notin\RR$. In particular, the assumption of Proposition~\ref{prop:general example} is satisfied.

By Propositions~\ref{prop:C_one}--\ref{prop:general example}, one may assume, by possibly removing a null set from $\Omega$, that for all $\bsp \in \Omega$, the channel $\Phi\in\cC_{d,\sigma}$ is irreducible, verifies \QDB{} with an anti-unitary $J$ such that $J^2=\one$, and satisfies the conclusions of Proposition~\ref{prop:general example}. 

Fix now any $\bsp\in \Omega$, and assume that for some unitary operator $G$, one has $G^\ast\Phi^\rho(GXG^\ast)G=\Phi(X)$ for all $X\in\cB(\cH)$. Since $\Phi$ satisfies the \QDB{} condition with $J$, one easily shows that it must be covariant w.r.t.\;the anti-unitary operator $\Gamma=J^\ast G$. By Proposition~\ref{prop:general example}(ii), this implies that $C_{0,1}C_{1,2}C_{2,0}\in\RR$, which is a contradiction.
\QED

\noindent{\bf Proof of Proposition~\ref{prop:unitary no 1}}.
Consider the set $\cC_{d,\sigma}$ with even $d\geq 6$, $a_0=d-1$, $\hat d=d/2$, and unitary $J$.
Set
$$
\eta_a=\begin{cases}
1&\text{if $a$ is even};\\[4pt]
\i&\text{otherwise,}
\end{cases}
$$ 
so that $\eta_{\sigma(a-1)}\eta_a=(-1)^a$ for all $a\in\ZZd$; see Figure~\ref{fig:d6_unitary_no1}. Once again, we start by showing that the assumption of Proposition~\ref{prop:general example} is satisfied. For this, consider  $p_0=0=1-p_{\sigma(0)}$, and $p_a=2^{-a}=1-p_{\sigma(a)}$ for $a\in\llbracket1,\hat d-1\rrbracket$, 
so that the only equality between two distinct entries is $p_1=p_{d-2}=1/2$. 
Direct computation leads to
$$
C_{a,a+1}=\sqrt{p_{a-1}p_a}-\sqrt{(1-p_{a-1})(1-p_a)}
$$
for $a\in\ZZd$.
Thus, $C_{a,a+1}=0$ iff 
$a\in\{0,\hat d\}$. It follows that 
$$
|C_{r,1+r}|>0=|C_{0,1}|
$$
for $r\in\ZZd\setminus\{0,\hat d\}$. For $r=\hat d$, we now argue that
$$
|C_{\hat d,\hat d+2}|\neq |C_{0,2}|.
$$
Indeed, for $\hat d=3,4$, the left-hand side is respectively
$
(1+\sqrt3)/(2\sqrt2)$ and $(\sqrt3+\sqrt7)/(4\sqrt2)$, 
and both values are strictly larger than $2^{-1/2}=|C_{0,2}|$. For $\hat d\geq5$,
$$
|C_{\hat d,\hat d+2}|<2^{2-\frac{\hat d}{2}}\leq2^{-\frac12}=|C_{0,2}|.
$$
The assumption of Proposition~\ref{prop:general example} is thus established. In view of this and Propositions~\ref{prop:C_one}~and~\ref{prop:C_two}, for almost all   $\bsp\in\fP_{d,\sigma}$, the channel $\Phi\in\cC_{d,\sigma}$ is irreducible of period $d$, satisfies \QDB{} with unitary $J$ such that $\Phi(J^2)=-J^2$ and satisfies the conclusions of Proposition~\ref{prop:general example}. We now fix any such $\bsp$.

Assume that $\Phi$ also satisfies the \QDB{} condition w.r.t.\;some unitary $G$ such that $\Phi(G^2)=G^2$. Since $\Phi$ is irreducible that implies $G^2$ is proportional to the identity.
The \QDB{} condition for $J$ and $G$ implies that $\Phi$ is covariant 
w.r.t.\;the unitary $\Gamma=J^\ast G$. By Proposition~\ref{prop:general example}(i), one has $\Gamma e_a=z\xi_d^{na}e_a$ for all $a\in\ZZd$, and some
$z\in\TT$ and $n\in\ZZ$. It follows that $Ge_a=J\Gamma e_a=z\xi_d^{na}\eta_ae_{\sigma(a-1)}$,
which leads to the contradiction $G^2e_a=(-1)^az^2e_a$ since there exists $\zeta\in\TT$ such that, for every $a\in\ZZd$,
$$
\zeta e_a=G^2e_a=Gz\xi_d^{na}\eta_ae_{\sigma(a-1)}
=z^2\eta_a\eta_{\sigma(a-1)}\xi_d^{n(a_0+1)}e_a=(-1)^az^2e_a.
$$
\QED

\section{Proofs of main results}
\label{sec:proof}

\subsection{Proof of Proposition~\ref{prop:conv_ep}}
\label{sec:proof-TRI-ep}

\noindent{\bf Part~(i)} We start with the Donsker--Varadhan variational formula
for the relative entropy of two probability measures $\PP$, $\QQ$ on $\Omega$,
\cite[Theorem~2.1]{Donsker1983},
\beq
\Ent(\PP|\QQ)=\sup_{f\in C_b(\Omega)}\left(
\int f\d\PP-\log\int \e^f\d\QQ\right),
\label{eq:DVform}
\eeq
where the supremum is taken over the set $C_b(\Omega)$ of all bounded continuous
real functions on $\Omega$. In particular,
$$
\Ep(\PP_{n+m},\theta)=\Ent(\PP_{n+m}|\wP_{n+m})=\sup_{f\in C_b(\Omega_{n+m})}\left(\int f\d\PP_{n+m}
-\log\int \e^f\d\wP_{n+m}\right),
$$
and restricting this supremum to functions $f=g+h\circ\phi^n$
where $g\in C_b(\Omega_n)$ and $h\in C_b(\Omega_m)$, we obtain the lower bound
$$
\Ep(\PP_{n+m},\theta)\geq\sup_{\twol{g\in C_b(\Omega_n)}{h\in C_b(\Omega_m)}}
\left(\int \left(g+h\circ\phi^n\right)\d\PP_{n+m}
-\log\int\e^g\e^{h\circ\phi^n}\d\wP_{n+m}\right).
$$
Since $\PP$ is $\phi$-invariant, we have
$$
\int\left(g+h\circ\phi^n\right)\d\PP_{n+m}=\int g\d\PP_n+\int h\d\PP_m,
$$
and by Assumption~\ER, which is also satisfied by $\wP$,
$$
\log\int\e^g\e^{h\circ\phi^n}\d\wP_{n+m}\leq\log\int\e^g\d\wP_{n}
+\log\int\e^h\d\wP_{m}+\log C.
$$
It follows that
$$
\Ep(\PP_{n+m},\theta)\ge\Ep(\PP_n,\theta)+\Ep(\PP_m,\theta)-\log C,
$$
which shows that the sequence $(\log C-\Ep(\PP_n,\theta))_{n\in\NN}$ is
subadditive. Fekete's lemma~\cite[Part~I, Chapter~3]{Polya1978} yields
$$
\ep(\PP,\theta)=\lim_{n\to\infty}\frac{\Ep(\PP_n,\theta)}n
=\sup_{n>0}\frac{\Ep(\PP_n,\theta)-\log C}n.
$$

\medskip\noindent
{\bf Part~(ii)} Obviously, if $\PP=\wP$, then $\Ep(\PP_n,\theta)=\Ent(\PP_n|\wP_n)=0$
for all $n$ and hence $\ep(\PP,\theta)=0$. Conversely, if $\ep(\PP,\theta)=0$,
then the above variational expression gives that, for all $n$,
$$
\Ep(\PP_n,\theta)\leq\log C.
$$
Invoking again the Donsker--Varadhan variational formula~\eqref{eq:DVform},
we observe that, with respect to the weak topology, relative entropy is a
jointly lower semi-continuous function of its two arguments. Since,
for an arbitrary probability measure $\QQ$ on $\Omega$, we have
$\PP_n\otimes\QQ\rightharpoonup\PP$ as $n\to\infty$, it follows that
$$
\Ent(\PP|\wP)\leq\liminf_{n\to\infty}\;\Ent(\PP_n\otimes\QQ|\wP_n\otimes\QQ)
=\liminf_{n\to\infty}\;\Ep(\PP_n,\theta)\leq \log C<\infty.
$$
From this finite bound, we deduce that $\PP$ is absolutely continuous
with respect to $\wP$. Since $\PP$ is assumed to be $\phi$-ergodic,
so is $\wP$ by the following Lemma, and since any two distinct $\phi$-ergodic measures
are mutually singular \cite[Theorem 6.10]{Walters1982}, we conclude that $\PP=\wP$.
\QED

\medskip
\bel\label{lem:ergodic}
$\PP\in\cP_\phi(\Omega)$ is $\phi$-ergodic iff\, $\wP$ is $\phi$-ergodic.
\eel

\proof Since the cylinder sets form a semi-algebra generating $\cF$, $\PP$ is
$\phi$-ergodic iff, for any cylinder sets $A,B\in\cF$,
\beq
\lim_{n\to\infty}\frac1n\sum_{k=0}^{n-1}\PP(A\cap \phi^{-k}(B))=\PP(A)\PP(B),
\label{eq:ergorel}
\eeq
and similarly for $\wP$, see e.g.~\cite[Theorem~1.17(i)]{Walters1982}.
Consider the cylinders
\begin{eqnarray*}
A=[A_1,\ldots,A_l],&\qquad& B=[B_1,\ldots,B_m],\\[4pt]
\widehat{A}=[\widehat A_l,\ldots,\widehat A_1],&\qquad& \widehat{B}=[\widehat B_m,\ldots,\widehat B_1],
\end{eqnarray*}
where $\widehat A_i=\theta(A_i)$ and $\widehat B_i=\theta(B_i)$. For $k\ge l$, one has
\beq
A\cap \phi^{-k}(B)=[A_1,\ldots,A_l,\underbrace{\bA,\ldots,\bA}_{k-l},B_1,\ldots,B_m],
\label{eq:cylcomp}
\eeq
and hence,
$$
\wP(A\cap \phi^{-k}(B))
=\PP([\widehat B_m,\ldots,\widehat B_1,\bA,\ldots,\bA,
\widehat A_l,\ldots,\widehat A_1]).
$$
Since
$$
[\widehat B_m,\ldots,\widehat B_1,\bA,\ldots,\bA,\widehat A_l,\ldots,\widehat A_1]
=[\widehat B_m,\ldots,\widehat B_1]\cap\phi^{-(k+m-l)}([\widehat A_l,\ldots,\widehat A_1]),
$$
we conclude that
$$
\wP(A\cap \phi^{-k}(B))=\PP(\widehat{B}\cap\phi^{-(k+m-l)}(\widehat{A})),
$$
and therefore
$$
\sum_{k=l}^{n-1}\wP(A\cap \phi^{-k}(B))
=\sum_{k=0}^{n-l-1}\PP(\widehat{B}\cap\phi^{-(k+m)}(\widehat{A})).
$$
It follows that if $\PP$ is $\phi$-ergodic, then
$$
\lim_{n\to\infty}\frac1n\sum_{k=0}^{n-1}\wP(A\cap \phi^{-k}(B))
=\PP(\widehat B)\PP(\widehat A)=\wP(A)\wP(B),
$$
which shows that $\wP$ is $\phi$-ergodic. Exchanging the roles of $\PP$ and $\wP$
establishes the opposite implication.\QED

\subsection{Proof of Proposition~\ref{prop:dilation}}
\label{sec:proofDilation}

\noindent{\bf Part~(i)} is a combination of~\cite[Proposition~7.5 and Theorem~7.5]{Busch2016}.

\noindent{\bf Part~(ii)}
Theorem~7.11 of~\cite{Busch2016} gives a Stinespring dilation together with a
projection-valued measure representing $\cJ$. The point here is to obtain a
representing POVM for any prescribed Stinespring dilation of $\Phi$.
Let $(\cE,V)$ be a Stinespring dilation of $\Phi\in\CPU(\cH)$, and $\cJ$ be a $(\Phi,\bA)$-instrument.
Invoking~\cite[Theorem~7.11]{Busch2016}, there exists a Hilbert space $\cK$, a POVM $P:\cA\to\cB_+(\cK)$
(which can be taken to be projection valued) and a linear isometry $W:\cH\to\cH\otimes\cK$
such that
$$
\cJ(A)(X)=W^\ast(X\otimes P(A))W
$$
for all $A\in\cA$ and $X\in\BH$. Denote by $\cG$ the linear span of the set
$$
G=\{(X\otimes\one_\cE)Vx\mid X\in\BH,x\in\cH\}\subset\cH\otimes\cE.
$$
Then, the formula
$$
T\sum_\alpha (X_\alpha\otimes\one_\cE)Vx_\alpha
=\sum_\alpha(X_\alpha\otimes\one_\cK)Wx_\alpha
$$
where the sums are over a finite set of indices, $X_\alpha\in\BH$ and $x_\alpha\in\cH$, defines
a linear map $T:\cG\to\cH\otimes\cK$. Moreover, since
\begin{align*}
\|\sum_\alpha (X_\alpha\otimes\one_\cK)Wx_\alpha\|^2
&=\sum_{\alpha,\beta}\bra x_\beta,W^\ast(X_\beta^\ast X_\alpha\otimes\one_\cK)Wx_\alpha\ket\\
&=\sum_{\alpha,\beta}\bra x_\beta,\Phi(X_\beta^\ast X_\alpha)x_\alpha\ket\\
&=\sum_{\alpha,\beta}\bra x_\beta,V^\ast(X_\beta^\ast X_\alpha\otimes\one_\cE)Vx_\alpha\ket\\
&=\|\sum_\alpha (X_\alpha\otimes\one_\cE)Vx_\alpha\|^2,
\end{align*}
$T$ is isometric. Introducing a basis $(e_1,\ldots,e_n)$ of $\cH$, we can write
$$
Ve_i=\sum_{j=1}^ne_j\otimes v_{ji}
$$
where $v_{ji}\in\cE$. Thus, defining the linear maps $V_j:\cH\to\cE$ by $V_je_i=v_{ji}$, one has
$$
Vx=\sum_{j=1}^ne_j\otimes V_jx,\qquad
V^\ast(e_j\otimes y)=V_j^\ast y,\qquad
\sum_{j=1}^nV_j^\ast V_j=\one_\cH,
$$
and similarly, for linear maps $W_j:\cH\to\cK$,
$$
Wx=\sum_{j=1}^ne_j\otimes W_jx,\qquad
W^\ast(e_j\otimes y)=W_j^\ast y,\qquad
\sum_{j=1}^nW_j^\ast W_j=\one_\cH.
$$
Let $E_{ij}\in\BH$ denote matrix units, so that $E_{ij}e_k=\delta_{k,j}e_i$. It follows that
\begin{align*}
(E_{ij}\otimes\one_\cE)Vx&=\sum_k(E_{ij}e_k\otimes V_kx)=e_i\otimes V_jx,\\
(E_{ij}\otimes\one_\cK)Wx&=\sum_k(E_{ij}e_k\otimes W_kx)=e_i\otimes W_jx,
\end{align*}
from which we conclude that $\Ran(V)\subset\cG=\cH\otimes\hat\cE$ and $\Ran(T)=\cH\otimes\hat\cK$, where
$$
\hat\cE=\sum_{j=1}^n\Ran(V_j)\subset\cE,\qquad
\hat\cK=\sum_{j=1}^n\Ran(W_j)\subset\cK,
$$
are finite-dimensional subspaces.
It follows that $T(e_i\otimes V_jx)=e_i\otimes W_jx$, from which we deduce that
$T=\one_\cH\otimes S$ for some linear isometry $S:\hat\cE\to\cK$ such that $\Ran(S)=\hat\cK$
and $SV_j=W_j$. Thus, extending $S$ by zero on $\hat\cE^\perp$,
$$
\cJ(A)(X)=W^\ast(X\otimes P(A))W=V^\ast(X\otimes S^\ast P(A)S)V.
$$
Let $P_{\hat\cE}$ be the orthogonal projection onto $\hat\cE$, and let
$\nu$ be an arbitrary probability measure on $(\bA,\cA)$. Then
$$
M(A)=S^\ast P(A)S+\nu(A)(\one_\cE-P_{\hat\cE})
$$
defines a POVM on $\cE$, since $S^\ast S=P_{\hat\cE}$ and hence
$M(\bA)=\one_\cE$. Moreover, the added term does not contribute to
$V^\ast(X\otimes M(A))V$, because
$\Ran(V)\subset\cH\otimes\hat\cE$. Thus, $M$ is a
$({\cJ},\cE,V)$-POVM.

\subsection{Proof of Proposition~\ref{prop:prop channel}}
\label{sec:proof prop channel}

Parts~(i),~(ii),~(iii) and~(iv) follow from~\cite[Lemma~4.1]{Evans1987}, \cite[Theorem~4.2]{Evans1987} 
and the discussion following it.

Part~(v) was first proved in~\cite[Section~5]{Kuemmerer2003}. We provide a similar proof
adapted to our definitions. Invoking again~\cite[Theorem~1.17(i)]{Walters1982},
it suffices to show that~\eqref{eq:ergorel} holds for any cylinder sets $A=[A_1,\ldots,A_l]$
and $B=[B_1,\ldots,B_m]$. Recalling~\eqref{eq:cylcomp} and using $\cJ(\bA)=\Phi$, we have
for $k\geq l$,
$$
\PP(A\cap \phi^{-k}(B))=\bra\one,\cJ(A_1)\cdots\cJ(A_l)\Phi^{k-l}\cJ(B_1)\cdots\cJ(B_m)\one\ket_\KMS.
$$
Moreover, from~(i) and~(ii), we infer that for any $X\in\BH$,
$$
\lim_{n\to\infty}\frac1n\sum_{k=l}^{n-1}\Phi^{k-l}(X)=\one\bra\one,X\ket_\KMS.
$$
It follows that
\begin{align*}
	\lim_{n\to\infty}\frac1n\sum_{k=0}^{n-1}\PP(A\cap \phi^{-k}(B))
	&=\lim_{n\to\infty}\frac1n\sum_{k=l}^{n-1}\PP(A\cap \phi^{-k}(B))\\
	&=\bra\one,\cJ(A_1)\cdots\cJ(A_l)\one\ket_\KMS\bra\one,\cJ(B_1)\cdots\cJ(B_m)\one\ket_\KMS\\[6pt]
	&=\PP(A)\PP(B),
\end{align*}
which completes the proof.

\subsection{Proof of Proposition~\ref{prop:possible values of c}}
\label{sec:proof possible values of c}

\begin{lemma}\label{lem:J permut}
	Let $J$ be a $(\Phi,\rho)$-admissible (anti-)unitary operator such that $\widehat{\Phi}=\Phi$
	for some irreducible $\Phi$ of period $p$. Let $(P_\alpha)_{\alpha\in\TT_p}$ be a maximal cycle of $\Phi$. 
	Then, there exists $\beta\in\TT_p$ such that
	$j(P_\alpha)=P_{\sigma(\alpha)}$, where $\sigma:\TT_p\to\TT_p$ denotes the involution defined by 
	$\sigma(\alpha)=\beta\alpha^{-1}$.
\end{lemma}

\proof
By Proposition~\ref{prop:prop channel}(iv), $[\rho,P_\alpha]=0$, and hence, for all $X\in\cB(\cH)$,
\begin{align*}
\bra\Phi^\rho(P_\alpha),X\ket_\KMS&=\bra P_\alpha,\Phi(X)\ket_\KMS=\tr(\rho^{1/2}P_\alpha\rho^{1/2}\Phi(X))
\\
&=\tr(\rho P_\alpha\Phi(X)P_\alpha)=\bra\rho,P_\alpha\Phi(X)P_\alpha\ket_\HS.  
\end{align*}
Since $\Phi(P_\alpha P_\alpha^\ast)=\Phi(P_\alpha)=P_{\tau(\alpha)}=P_{\tau(\alpha)}P_{\tau(\alpha)}^\ast=\Phi(P_\alpha)\Phi(P_\alpha^\ast)$, 
it follows from~\cite[Theorem~8.6]{Alicki2001} that
$\Phi(P_\alpha XP_\alpha)=P_{\tau(\alpha)}\Phi(X)P_{\tau(\alpha)}$. We deduce that
\begin{align*}
	\bra\Phi^\rho(P_\alpha),X\ket_\KMS&=\bra\rho,\Phi(P_{\tau^{-1}(\alpha)}XP_{\tau^{-1}(\alpha)})\ket_\HS
=\bra\Phi^\ast(\rho),P_{\tau^{-1}(\alpha)}XP_{\tau^{-1}(\alpha)}\ket_\HS\\[4pt]	&
	=\bra\rho,P_{\tau^{-1}(\alpha)}XP_{\tau^{-1}(\alpha)}\ket_\HS=\tr(P_{\tau^{-1}(\alpha)}\rho P_{\tau^{-1}(\alpha)}X)=\tr(\rho^{1/2}P_{\tau^{-1}(\alpha)}\rho^{1/2}X)\\[4pt]
	&=\bra P_{\tau^{-1}(\alpha)},X\ket_\KMS,
\end{align*}
so that $\Phi^\rho(P_\alpha)=P_{\tau^{-1}(\alpha)}$. Since $j^{-1}\circ\Phi^\rho=\Phi\circ j^{-1}$, it
follows that $\Phi\circ j^{-1}(P_\alpha)=j^{-1}(P_{\tau^{-1}(\alpha)})$. Thus, setting $Q_\alpha=j^{-1}(P_{\alpha^{-1}})$ 
yields a maximal cycle $(Q_\alpha)_{\alpha\in\TT_p}$ of $\Phi$. By uniqueness, there exists $\beta\in\TT_p$
such that $Q_\alpha=P_{\beta\alpha}$, and so $j(P_\alpha)=P_{\beta\alpha^{-1}}$.\QED

\medskip
We now proceed with the proof of Proposition~\ref{prop:possible values of c}.
Under irreducibility, the two identities in the statement imply $j(\rho)=\rho$, so $J$ is $(\Phi,\rho)$-admissible.

\noindent$(i)$ $J$ being unitary, we have 
$\Phi^\rho(J^2)=J\widehat{\Phi}(J^*J^2J)J^*=J\widehat{\Phi}(J^2)J^*=J\Phi(J^2)J^*=\eta J^2$. 
Thus, since $J\rho=\rho J$,
$$
\overline\eta=\langle\eta J^2,J^2\rangle_\KMS=\langle\Phi^\rho(J^2),J^2\rangle_\KMS
=\langle J^2,\Phi(J^2)\rangle_\KMS=\langle J^2,\eta J^2\rangle_\KMS=\eta,
$$
and we conclude that $\eta\in\RR\cap\TT$, as required.

\medskip\noindent$(ii)$
We use the notation of Proposition~\ref{prop:prop channel} and observe that Lemma~\ref{lem:J permut}
gives $j(U)=j^{-1}(U)=\beta^{-1}U$. Since   
$\Phi(U^{n\ast}U^n)=\Phi(\one)=\one=\Phi(U^{n\ast})\Phi(U^n)$, we deduce 
from~\cite[Theorem~8.6]{Alicki2001} that for $X\in\cB(\cH)$ and $n,m\in\ZZ$,
$U^{n\ast}\Phi(U^nXU^{m\ast})U^m=\xi_p^{n-m}\Phi(X)$. For $\alpha\in\widetilde\TT_p$,
set $J_\alpha=U^nJ$, $n$ being such that $\xi_p^{2n}=\alpha$. Since $U\rho=\rho U$ and
$J\rho=\rho J$, one also has $J_\alpha\rho=\rho J_\alpha$. We derive
$$
J_\alpha=Jj^{-1}(U^n)=J\beta^{-n}U^n=\beta^nJU^n,
$$
and hence
$$
J_\alpha^\ast\Phi^\rho(J_\alpha XJ_\alpha^\ast)J_\alpha=U^{n\ast}\Phi(U^nXU^{n\ast})U^n=\Phi(X).
$$ 
Moreover, writing $J_\alpha^2=\beta^{-n}U^nJ^2U^n$, we get
$$
\Phi(J_\alpha^2)=\beta^{-n}\Phi(U^nJ^2U^n)=\beta^{-n}\xi_p^{2n}U^{n}\Phi(J^2)U^n=\eta\alpha J_\alpha^2.
$$
\QED

\subsection{Preliminaries on channels and instruments}
\label{sec:prelimChannels}
\bel\label{lem:proof ER}
Let $\Phi$ be a quantum channel with faithful invariant state $\rho$, and
let $\cJ$ be a $(\Phi,\bA)$ instrument. Then, the $\rho$-statistics of $\cJ$
satisfies Assumption~\ER, with $C=\|\rho^{-1}\|$.
\eel

\proof
The proof is an adaptation of~\cite[Lemma~3.4]{Benoist2017c}. Since cylinder
sets form a semi-algebra generating $\cF$, it suffices to
show that there exists a constant $C$ such that the inequality in~\ER{}
holds for two arbitrary cylinder sets $A=[A_1,\ldots,A_n]$ and $B=[B_1,\ldots,B_m]$.
By \eqref{eq:Pdef} and \eqref{eq:Holder}, one has
\begin{align*}
\PP(A\cap\phi^{-n}(B))&=\PP([A_1,\ldots,A_n,B_1,\ldots,B_m])\\
&=\bra\one,\cJ(A_1)\cdots\cJ(A_n)\cJ(B_1)\cdots\cJ(B_m)\one\ket_\KMS\\
&=\bra(\cJ(A_1)\cdots\cJ(A_n))^\rho\one,\cJ(B_1)\cdots\cJ(B_m)\one\ket_\KMS\\
&\le\|(\cJ(A_1)\cdots\cJ(A_n))^\rho\one\|\bra\one,\cJ(B_1)\cdots\cJ(B_m)\one\ket_\KMS\\
&=\|\rho^{-1/2}(\cJ(A_1)\cdots\cJ(A_n))^\ast(\rho)\rho^{-1/2}\|\,\PP(B)\\
&\le\|\rho^{-1/2}\|^2\|(\cJ(A_1)\cdots\cJ(A_n))^\ast(\rho)\|\,\PP(B).
\end{align*}
Setting $C=\|\rho^{-1/2}\|^2=\|\rho^{-1}\|=1/\min\sp(\rho)$, and using the fact that
$T\ge0$ implies $\|T\|\le\tr(T)$, we further get
\begin{align*}
\PP(A\cap\phi^{-n}(B))
&\le C\,\tr((\cJ(A_1)\cdots\cJ(A_n))^\ast(\rho))\,\PP(B)\\
&=C\,\bra(\cJ(A_1)\cdots\cJ(A_n))^\ast\rho,\one\ket\PP(B)\\
&=C\,\bra\rho,\cJ(A_1)\cdots\cJ(A_n)\one\ket\PP(B)\\
&=C\,\bra\one,\cJ(A_1)\cdots\cJ(A_n)\one\ket_\KMS\PP(B)\\
&=C\,\PP(A)\PP(B),
\end{align*}
which completes the proof.\QED

\subsection{Preliminaries on POVMs}
\label{sec:prelimPOVM}

We note that given two CP maps $\Phi_1,\Phi_2:\BH\to\BH$, with Stinespring
dilations $(\cE_1,V_1)$ and $(\cE_2,V_2)$, one can always assume that $\cE_1=\cE_2$.
Indeed, setting $\cE=\cE_1\oplus\cE_2$, and replacing $V_1,V_2$ with the maps
$$
\widetilde{V}_1:\cH\ni x\mapsto V_1x\oplus0,\qquad
\widetilde{V}_2:\cH\ni x\mapsto 0\oplus V_2x,
$$
yields the two dilations $(\cE,\widetilde{V}_1)$ and $(\cE,\widetilde{V}_2)$.

The following basic facts about IC-POVMs will be useful.
\bel\label{lem:IC}
Let $\cE$ be a separable Hilbert space.
\begin{enumerate}[(i)]
\item There exists a measurable space $(\bA,\cA)$ and an IC-POVM $M:\cA\to\cB_+(\cE)$.
\item If $(\bA,\cA)$ is a measurable space and $M:\cA\to\cB_+(\cE)$ an IC-POVM, then the
linear span of $M(\cA)$ is weak*-dense in $\cB(\cE)$.
\end{enumerate}
\eel
\begin{remarks} {\bf (i)} Although (i) is known, see Example 3 in~\cite{Busch1995}, for the reader's convenience we provide a proof along different lines. 

\noindent{\bf (ii)} If $\dim \cE=d$, then $|\bA|\geq  d^2$, and one can construct the so-called minimal  IC-POVM's with $|\bA|=d^2$, see~\cite{Caves2002} 
and~\cite[Example~3.50]{Heinosaari2011}. In the  construction presented below $|\bA|=4d(d-1)$. 

\end{remarks}
\proof (i) Let $(e_l)_{l\in\Lambda}$ and $(u,d)$ be
orthonormal bases of $\cE$ and $\CC^2$ respectively. Let $P=[p_{kl}]_{k,l\in\Lambda}$ be
a bistochastic matrix such that
$$
p_{kl}\begin{cases}
=0&\text{if } k=l;\\
>0&\text{otherwise.}
\end{cases}
$$
If $\dim(\cE)=n<\infty$\footnote{We exclude the trivial case $\dim(\cE)=1$.}, one can set all the non-zero entries of $P$ to $1/(n-1)$. Otherwise, take $P=P^T$,
with the first line
$$
0,2^{-1},2^{-2},2^{-3},\ldots,
$$
the second line
$$
2^{-1},0,2^{-2},2^{-3},\ldots
$$

and, for $k\ge3$,  the $k^\mathrm{th}$ line
$$
2^{-(k-1)},2^{-(k-1)},2^{-(k-2)},\ldots,2^{-3},2^{-2},0,2^{-2},2^{-3},\ldots
$$
For
$(k,l)\in\Lambda^2$, with $k\neq l$, define a partial isometry $J_{kl}:\CC^2\to\cE$ by
$$
J_{kl}=e_k\otimes u^\ast+e_l\otimes d^\ast.
$$
Finally, let $\bA_0$ be any set of at least four elements, and let $N:2^{\bA_0}\to\cB_+(\CC^2)$ be an IC-POVM.\footnote{Such a POVM is easy to construct.}
Set
$\bA=\{(k,l,a)\in\Lambda\times\Lambda\times\bA_0\mid k\neq l\}$, $\cA=2^\bA$ and
$$
\cA\ni A\mapsto M(A)=\tfrac12\sum_{(k,l,a)\in A}p_{kl}J_{kl}N(\{a\})J_{kl}^\ast.
$$
By positivity of the summands, bistochasticity of $P$, and
$N(\bA_0)=\one_{\CC^2}$, the series defining $M(A)$ converges in the
weak* topology for every $A\in\cA$, the resulting map is weak*
$\sigma$-additive, and $M(\bA)=\one_\cE$. Thus, $M$ is a POVM. We now
show that it is informationally complete.

Indeed, for any $\ell\in\cT(\cE)$ and $A_0\in2^{\bA_0}$,
$$
\bra\ell,M(\{k\}\times\{l\}\times A_0)\ket_{\cE}=\tfrac12p_{kl}\bra\ell_{kl},N(A_0)\ket_{\CC^2},\qquad
\ell_{kl}=\begin{bmatrix}\bra e_k,\ell e_k\ket&\bra e_k,\ell e_l\ket\\
\bra e_l,\ell e_k\ket&\bra e_l,\ell e_l\ket
\end{bmatrix}.
$$
Thus, if $\bra\ell,M(A)\ket_\cE=0$ for all $A\in\cA$, then $\bra\ell_{kl},N(A_0)\ket_{\CC^2}=0$
for all $A_0\in2^{\bA_0}$ and all $k,l\in\Lambda$. Since $N$ is informationally complete, \cite[Proposition~18.1]{Busch2016} implies
$\ell_{kl}=0$ for all $k,l\in\Lambda$, and hence $\ell=0$.

\medskip\noindent(ii) This is a direct consequence of \cite[Proposition~18.1]{Busch2016}.
\QED

\medskip
The next result shows that informational completeness of POVMs is stable under tensor product.
\bel\label{lem:guillaumetell}
If, for $i\in\{1,2\}$, $M_i:\cA_i\to\cB_+(\cE_i)$ are IC-POVMs, then the map $M:\cA_1\otimes\cA_2\longrightarrow\cB_+(\cE_1\otimes\cE_2)$ defined, for any $A_1\in\cA_1$ and $A_2\in\cA_2$, by
$$
M(A_1\times A_2)=M_1(A_1)\otimes M_2(A_2)
$$
is an IC-POVM.
\eel
\proof
Let $\ell$ be a normal linear functional on $\cB(\cE_1\otimes\cE_2)$ such that
$$
\ell\circ M(A_1\times A_2)=0
$$
for all $(A_1\times A_2)\in\cA_1\times\cA_2$. Following \cite[Proposition~18.1]{Busch2016}, we have to show that $\ell=0$.
For $X\in\cB(\cE_1)$, denote by $\varphi_X$ the normal linear functional defined on $\cB(\cE_2)$
by $\varphi_X(Y)=\ell(X\otimes Y)$. By Lemma~\ref{lem:IC}(ii), it follows from
$$
0=\ell\circ M(A_1\times A_2)=\varphi_{M_1(A_1)}(M_2(A_2))
$$
that $\varphi_{M_1(A_1)}=0$ for all $A_1\in\cA_1$. For $Y\in\cB(\cE_2)$ consider now the normal
linear functional defined on $\cB(\cE_1)$ by $\psi_Y(X)=\ell(X\otimes Y)$. We have
$$
0=\varphi_{M_1(A_1)}(Y)=\ell(M_1(A_1)\otimes Y)=\psi_Y(M_1(A_1)),
$$
and, again by Lemma~\ref{lem:IC}(ii), we conclude that $\psi_Y=0$ for all $Y\in\cB(\cE_2)$.
Consequently,
$$
\ell(X\otimes Y)=\psi_{Y}(X)=0
$$
for all $X\in\cB(\cE_1)$ and $Y\in\cB(\cE_2)$. Since
finite sums of tensor products are ultraweakly dense in $\cB(\cE_1\otimes\cE_2)$, we obtain
$\ell=0$, as required.\QED

\subsection{Relating the Stinespring dilations of $\Phi$ and $\widehat{\Phi}$}
\label{sec:Stinespring}

In this section we consider a quantum channel $\Phi\in\CPU(\cH)$ with a faithful
invariant state $\rho$. We also fix a $(\Phi,\rho)$-admissible unitary/anti-unitary
map $J:\cH\to\cH$ and the associated $\ast$-morphism $j:\BH\to\BH$. 
$\cH^\ast$ denotes the anti-dual space of the Hilbert space $\cH$, \ie the set
of linear functionals $y^\ast:\cH\ni x\mapsto\bra y,x\ket_\cH$, with
the laws of addition and scalar multiplication
$$
y^\ast+\lambda z^\ast=(y+\overline\lambda z)^\ast.
$$
The sesquilinear map $\bra y^\ast,z^\ast\ket_{\cH^\ast}=\bra z,y\ket_\cH$
makes $\cH^\ast$ a complex Hilbert space. Given an operator $A:\cH\to\cH$,
we denote by the same symbol the operator defined on $\cH^\ast$ by
$Ax^\ast=(A^\ast x)^\ast$. One easily checks that the latter is linear/anti-linear whenever the former is.
Moreover, if $A$ acts unitarily/anti-unitarily on $\cH$, so does it on $\cH^\ast$.
We will use without further mention the canonical identification of $\cH\otimes\cH^\ast$
with $\BH$.\footnote{With Dirac's notation, $x\otimes y^\ast\in\cH\otimes\cH^\ast$ is identified
with the rank-one operator $|x\ket\bra y|\in\BH$.} Under the latter,
the map $J\otimes J^\ast$ acts on $\BH$ as $j$. Note that
with this identification $(x\otimes y^\ast)^\ast=y\otimes x^\ast$,
the map $X\mapsto X^\ast$ being an anti-unitary involution on $\BH$
w.r.t.\;both the Hilbert-Schmidt and the KMS inner products.

We shall need the following result, which is likely  well known. We provide a proof
since we lack a convenient  reference.

\bel\label{lem:unitimage}
Let $A$ and $B$ be operators on the Hilbert space $\cH$ such that $\|Ax\|=\|Bx\|$ for all $x\in\cH$. Then, there exists a partial isometry\footnote{The notation $J:\cV\rightleftarrows\cW$ means that $J$ is a partial isometry with initial/final space $\cV$/$\cW$.} $U:\overline{\Ran A}\rightleftarrows\overline{\Ran B}$ such that $B=UA$. If both $A$ and $B$ are
linear or anti-linear, then $U$ is linear. If one of them is linear and the other anti-linear, then $U$ is anti-linear.
\eel
\proof
Let $C$ be an arbitrary conjugation on $\cH$, \ie an anti-linear operator
such that $C=C^\ast=C^{-1}$. We set
$$
\hat A=\begin{cases}
A&\text{if }A\text{ is linear};\\
CA&\text{if }A\text{ is anti-linear},\\
\end{cases}
$$
and define $\hat B$ similarly. Then, both $\hat A$ and $\hat B$ are linear with polar decompositions
$$
\hat A=S|A|,\qquad \hat B=T|B|,
$$
where, by polarization,  $|A|=(A^\ast A)^{1/2}=(\hat{A}^\ast\hat A)^{1/2}=(\hat{B}^\ast\hat B)^{1/2}=(B^\ast B)^{1/2}=|B|$, while
$$
S:\overline{\Ran A^\ast}\to\overline{\Ran \hat A},\qquad
T:\overline{\Ran B^\ast}\to\overline{\Ran \hat B},
$$
are partial isometries. Since $\Ker A=\Ker B$ and hence $\overline{\Ran A^\ast}=\overline{\Ran B^\ast}$,
it follows that
$$
\hat B=TS^\ast \hat A=\hat U\hat A,
$$
where $\hat U:\overline{\Ran \hat A}\to\overline{\Ran \hat B}$ is a partial isometry.
Defining $U$ to be $\hat U$, $C\hat U$, $\hat U C$ or $C\hat U C$ depending on the nature of $A$ and $B$
yields the required partial isometry.\QED

\bed\label{def:PsiV}
Let $(\cE,V)$ be a Stinespring dilation of a quantum channel $\Phi$.
\begin{enumerate}[label=(\roman*)]
\item To $O\in\cB(\cE)$, we associate the following linear maps on $\BH$,
$$
\fS_{V,O}:X\mapsto V^*(X\otimes O)V,\qquad
\widehat{\fS}_{V,O}=j^{-1}\circ\fS_{V,O}^\rho\circ j.
$$
\item Let $\psi_V:\BH\to\cE$ be the linear map defined by the partial trace
$$
\psi_V(X)=\tr_\cH(V\rho^{\frac12}X),
$$
and set $\cE_V=\Ran\psi_V$.
\end{enumerate}
\eed

\noindent{\bf Remarks.}\begin{enumerate}[label=\arabic*.]
\item If $O\in\cB_+(\cE)$, then $\fS_{V,O}$ is completely positive. In particular,
for any Stinespring dilation $(\cE,\widehat{V})$ of $\widehat{\Phi}$,
$$
\fS_{V,\one_\cE}=\Phi,\qquad
\widehat{\fS}_{V,\one_\cE}=\widehat\Phi=\fS_{\widehat{V},\one_\cE}.
$$
The next lemma is an extension of the last relation.
\item For any POVM $M:\cA\to\cB_+(\cE)$, the map $\cA\ni A\mapsto \fS_{V,M(A)}$ is a $(\Phi,\bA)$-instrument.
\item Since $\BH$ is finite dimensional, so is $\cE_V\subset\cE$. Moreover,
for $(x,y^\ast)\in\cH\times\cH^\ast$, one has
$$
\psi_V(x\otimes y^\ast)=\tr_\cH(V\rho^{1/2}(x\otimes y^\ast))
=(y^\ast\otimes\one_\cE)V\rho^{1/2}x,
$$
and hence
\beq
\begin{aligned}
\bra\psi_V(x_1\otimes y_1^\ast),O\psi_V(x_2\otimes y_2^\ast)\ket_\cE
&=\bra(y_1^\ast\otimes\one_{\cE})V\rho^{1/2}x_1,
O(y_2^\ast\otimes\one_{\cE})V\rho^{1/2}x_2\ket_\cE\\
&=\bra x_1,
\rho^{1/2}V^\ast((y_1\otimes y_2^\ast)\otimes O)V\rho^{1/2}x_2\ket_\cH\\
&=\tr_\cH(\rho^{1/2}\fS_{V,O}(y_1\otimes y_2^\ast)\rho^{1/2}(x_2\otimes x_1^\ast))\\
&=\bra x_1\otimes x_2^*,\fS_{V,O}(y_1\otimes y_2^*)\ket_\KMS
\end{aligned}
\label{eq:KMStrick}
\eeq
for any $(x_1,y_1^\ast),(x_2,y_2^\ast)\in\cH\times\cH^\ast$ and $O\in\cB(\cE)$.
\item For the sake of generality, we will not restrict ourselves to minimal $\cE$
as is often done in the literature. We will only assume $\cE$ to be separable.
We note, however, that the finite-dimensional subspace $\cE_V$ will play the role
of a minimal subspace. Indeed, denoting by $P_V\in\cB(\cE)$ the orthogonal projection
onto $\cE_V$, we note that~\eqref{eq:KMStrick} implies that
$\fS_{V,O}=\fS_{V,P_VOP_V}$.
\end{enumerate}

\bel\label{lem:def C}
Let $(\cE,V)$ and $(\cE,\widehat{V})$ be Stinespring dilations of $\Phi$ and
$\widehat{\Phi}$ respectively, and assume that the map $J$ is anti-unitary
(respectively unitary). Then there exists a linear (respectively anti-linear)
partial isometry $U:\cE_{\widehat{V}}\rightleftarrows\cE_V$ such that
\beq
\widehat{\fS}_{V,O}=\fS_{\widehat{V},U^\ast OU}
\label{eq:toshowjjojj}
\eeq
for any $O\in\cB(\cE)$.
\eel
\proof
With $(x_1,y_1^\ast),(x_2,y_2^\ast)\in\cH\times\cH^\ast$, Relation~\eqref{eq:KMStrick}
and the fact that $j$ is anti-unitary (respectively unitary) w.r.t.\;the KMS inner
product yield
\begin{align*}
\bra\psi_V\circ j(x_1\otimes y_1^\ast),\psi_V\circ j(x_2\otimes y_2^\ast)\ket_\cE
=&\bra\psi_V(Jx_1\otimes(Jy_1)^\ast),\psi_V(Jx_2\otimes(Jy_2)^\ast)\ket_\cE\\
=&\bra j(x_1\otimes x_2^\ast),\Phi\circ j(y_1\otimes y_2^\ast)\ket_\KMS\\
=&\bra\Phi^\rho\circ j(x_1\otimes x_2^\ast),j(y_1\otimes y_2^\ast)\ket_\KMS\\
=&\bra y_1\otimes y_2^\ast,\widehat{\Phi}(x_1\otimes x_2^\ast)\ket_\KMS^\sharp\\
=&\bra\psi_{\widehat{V}}(y_1\otimes x_1^\ast),\psi_{\widehat{V}}(y_2\otimes x_2^\ast)\ket_\cE^\sharp\\
=&\bra\psi_{\widehat{V}}((x_1\otimes y_1^\ast)^\ast),\psi_{\widehat{V}}((x_2\otimes y_2^\ast)^\ast)\ket_\cE^\sharp,
\end{align*}
where $z^\sharp=z$ if $J$ is anti-unitary and $z^\sharp=\overline{z}$ if $J$ is unitary.
It follows that
$$
\|\psi_{\widehat{V}}(X)\|=\|\psi_V\circ j(X^\ast)\|
$$
for all $X\in\BH$. By Lemma~\ref{lem:unitimage}, there exists a linear (respectively anti-linear)
partial isometry $U:\Ran\psi_{\widehat{V}}\rightleftarrows\Ran\psi_V$ such that, for any $X\in\BH$,
$$
U\psi_{\widehat{V}}(X)=\psi_V\circ j(X^\ast).
$$

To prove~\eqref{eq:toshowjjojj}, we observe that Relation~\eqref{eq:KMStrick} further gives
\begin{align*}
\bra x_1\otimes x_2^\ast,\widehat{\fS}_{V,O}(y_1\otimes y_2^\ast)\ket_\KMS
&=\bra x_1\otimes x_2^\ast,j^\rho\circ\fS_{V,O}^\rho\circ j(y_1\otimes y_2^\ast)\ket_\KMS\\
&=\bra j(y_1\otimes y_2^\ast),\fS_{V,O}\circ j(x_1\otimes x_2^\ast)\ket_\KMS^\sharp\\
&=\bra\psi_V\circ j((x_1\otimes y_1^\ast)^\ast),O\psi_V\circ j((x_2\otimes y_2^\ast)^\ast)\ket_\cE\\
&=\bra U\psi_{\widehat{V}}(x_1\otimes y_1^\ast),OU\psi_{\widehat{V}}(x_2\otimes y_2^\ast)\ket_\cE\\
&=\bra\psi_{\widehat{V}}(x_1\otimes y_1^\ast),U^\ast OU\psi_{\widehat{V}}(x_2\otimes y_2^\ast)\ket_\cE\\
&=\bra x_1\otimes x_2^\ast,\fS_{\widehat{V},U^\ast OU}(y_1\otimes y_2^\ast)\ket_\KMS.
\end{align*}
\QED

It should be noted that the construction does not rely on Assumption~\QDB{} but only on the
fact that~\eqref{eq:defhatphi} defines an affine map $\Phi\mapsto\widehat{\Phi}$ on $\CPU(\cH)$.
As mentioned in Remark~\ref{rem:KMSpresQC}(ii), this is a distinctive property of the KMS inner product.

\subsection{Proof of Theorem~\ref{thm:main-exist-instrument}}
\label{sec:QDB implies IQDB}

The assumption~\IQDB{} implies $\Phi=\cJ(\bA)=\wJ(\bA)=\widehat{\Phi}$. Hence,
if~\IQDB{} holds for an informationally complete instrument, so does~\QDB.

For the opposite implication, assuming~\QDB, $\Phi$ and $\widehat{\Phi}$ share
the same Stinespring dilation.

\bel\label{lem:C involution}
Let $(\cE,V)$ be a Stinespring dilation of the quantum channel $\Phi$ satisfying~\QDB, with
the operator $J$ being anti-unitary (respectively unitary). Then there exists a linear
(respectively anti-linear) partial isometry $S:\cE_V\rightleftarrows\cE_V$ such that
\beq
\widehat{\fS}_{V,O}=\fS_{V,S^\ast OS}
\label{eq:sapho}
\eeq
for all $O\in\cB(\cE)$. Moreover, denoting by $P_V$ the orthogonal projection onto $\cE_V$, one
has $S^2=P_V$ if $S$ is linear and $S^2=\pm P_V$ if it is anti-linear.
\eel
\proof
Lemma~\ref{lem:def C} and the fact that $\widehat{V}=V$ give
$$
\widehat{\fS}_{V,O}=j^{-1}\circ\fS_{V,O}^\rho\circ j=\fS_{V,U^\ast OU},
$$
where $U$ is a linear (respectively anti-linear) partial isometry on $\cE_V$.
Taking the $\rho$-adjoint of these identities and using the fact that $j$ is $\rho$-anti-unitary
(respectively unitary, a consequence of the first condition in~\eqref{eq:Phi-rho-adm}) we derive
$$
\widehat{\fS}_{V,O}^\rho=j^{-1}\circ\fS_{V,O}\circ j=\fS_{V,U^\ast OU}^\rho.
$$
Invoking again Lemma~\ref{lem:def C}, conjugation with $j$ further yields
$$
j^{-1}\circ\widehat{\fS}_{V,O}^\rho\circ j=j^{-2}\circ\fS_{V,O}\circ j^2
=j^{-1}\circ\fS_{V,U^\ast OU}^\rho\circ j=\fS_{V,U^{\ast2}OU^2}.
$$
By the second condition in~\eqref{eq:Phi-rho-adm}, the completely positive unital map
$\Psi:\cB(\cH\otimes\cE)\to\BH$ defined by $\Psi(X\otimes O)=V^\ast(X\otimes O)V$ satisfies
$\Psi(J^2\otimes\one_\cE)=\Phi(J^2)=\eta J^2$. It follows from~\cite[Theorem~8.6]{Alicki2001} applied with $a=J^2\otimes\one_\cE$
and the previous identities that for all $X\in\BH$ and $O\in\cB(\cE)$,
$$
\Psi(X\otimes U^{\ast 2}OU^2)=\fS_{V,U^{\ast2}OU^2}(X)
=J^{2\ast}\Psi((J^2\otimes\one_\cE)(X\otimes O)(J^2\otimes\one_\cE)^\ast)J^2
=\Psi(X\otimes O).
$$
Thus, setting $D_O=U^{\ast 2}[O,U^2]$, we get $\Psi(X\otimes D_O)=0$.
With $X=x_1\otimes x_2^\ast$ and $Y=y_1\otimes y_2^\ast$, Relation~\eqref{eq:KMStrick}
allows us to write
$$
0=\bra Y,\Psi(X\otimes D_O)\ket_\KMS
=\bra U^2\psi_V(y_1\otimes x_1^\ast),[U^2,O]\psi_V(y_2\otimes x_2^\ast)\ket_\cE,
$$
from which we conclude that $P_V[U^2,O]P_V=0$. It follows that $U^2=\e^{2\i \varphi}P_V$ for
some $\varphi\in\RR$. If $U$ is anti-linear, Wigner's decomposition~\cite{Wigner1993a}
implies $U^2=\pm P_V$, and we can set $S=U$. In the opposite case, one takes
$S=\e^{-\i\varphi}U$.
\QED

\ber If $\cE_V\neq\cE$, an anti-linear $U$ with $U^2=-P_V$
may not have an anti-unitary extension to $\cE$ whose
square is $-\one_\cE$. This happens, for example, whenever
$\mathrm{codim}(\cE_V)=1$. 
\eer

\bel\label{lem:M general}
With the hypotheses and notations of Lemma~\ref{lem:C involution}, let $M$ be a
$\cB_+(\cE)$-valued POVM such that $P_VS^\ast M(\,\cdot\,)SP_V=P_VM\circ\theta(\,\cdot\,)P_V$
for some involution $\theta$. Then the $(\Phi,\bA)$-instrument $\cJ(\,\cdot\,)=\fS_{V,M(\,\cdot\,)}$
satisfies the~\IQDB{} condition with anti-unitary (respectively unitary) $J$.
\eel

\proof
First, the condition on $M$ ensures that $\theta$ is implementable within the Stinespring
dilation $(\cE_V,V)$, the required unitary (respectively anti-unitary) operator being given
by $S$. Next, using this condition, Relation~\eqref{eq:sapho} and the fact that $S^2=\pm P_V$,
we obtain
$$
\wJ(A)=\widehat{\fS}_{V,M\circ\theta(A)}=\widehat{\fS}_{V,S^\ast M(A)S}
=\fS_{V,S^{2\ast}M(A)S^2}=\fS_{V,M(A)}=\cJ(A),
$$
as claimed.\QED

In specific situations, there may be a natural choice of POVM $M$ satisfying the assumptions of
Lemma~\ref{lem:M general}. To address the general case, we now show that such POVMs always exist and, moreover, can be chosen to be informationally complete. Let $(\cE,V)$ be a Stinespring dilation of $\Phi$ such that $\cE_V=\cE$. Namely, it is a minimal dilation of $\Phi$. Let $(\bA',N)$ be any POVM on $\cE$.
Set $\bA=\{-1,1\}\times\bA'$, with the $\sigma$-algebra $\cA$ generated by the sets
$\{\pm1\}\times A'$, $A'\in\cA'$. Let $M$ be the POVM over $\cA$ defined by
$$M
(\{1\}\times A')=\frac12N(A'),\qquad
M(\{-1\}\times A')=\frac12S^\ast N(A')S,
$$
with $S$ as in Lemma~\ref{lem:C involution}. Remark that since $\cE_V=\cE$, $S$ is an isometry such that $S^2=\pm\one_\cE$. Then the assumptions of Lemma~\ref{lem:M general} hold with the involution
\begin{alignat*}{3}
\theta:&\quad \bA&\longrightarrow&\quad\bA\\
       &(\pm1,a)&\longmapsto& (\mp 1,a).
\end{alignat*}
Finally, notice that Lemma~\ref{lem:IC}(i) allows us to assume that $N$
is informationally complete. Then the same holds for $M$, which
proves Theorem~\ref{thm:main-exist-instrument}.

\subsection{Purely generated finitely correlated states}
\label{sec:PGFCS}

{\sl Purely generated finitely correlated states} (PGFCS) were
first introduced in~\cite{Fannes1992} as ground states of a family of gapped
local frustration-free spin Hamiltonians. They were then studied under the name
{\em matrix product states} (MPS) in the quantum information community as
approximations of ground states of local gapped spin Hamiltonians;
see~\cite{PerezGarcia2007,Verstraete2008} for introductions from this community
point of view.

Starting again with a Stinespring dilation $(\cE,V)$ of a channel $\Phi$ with
faithful invariant state $\rho$ we set $\cO=\cB(\cE)$.
With the notations
of the previous sections, 
the family $(\gamma^{(n)})_{n\in\NN}$ of linear forms on $\cO^{\otimes n}$ defined by
\beq
\cO^{\otimes n}\ni
A_1\otimes\ldots\otimes A_n\mapsto\gamma^{(n)}(A_1\otimes\cdots\otimes A_n)=\bra\one,\fS_{V,A_1}\cdots\fS_{V,A_n}\one\ket_\KMS
\label{eq:pgfcsDef}
\eeq
is consistent in the sense that $\gamma^{(n+1)}(A_1\otimes \cdots \otimes A_n\otimes \one_\cO)=\gamma^{(n)}(A_1\otimes \cdots \otimes A_n)$. Hence it
uniquely extends to the inductive limit state $\gamma$ on the inductive limit $C^\ast$-algebra $\cO^{\otimes\NN}$. In particular,
$$
\gamma(A\otimes\one_{\cO^{\otimes\NN}})=\gamma^{(n)}(A)
$$
for any $A\in\cO^{\otimes n}$; see~\cite[Section~1.23]{Sakai1971} for more details. As usual in this context, we will identify $\cO^{\otimes n}$
with a subalgebra of $\cO^{\otimes\NN}$ and write $A$ for $A\otimes\one_{\cO^{\otimes\NN}}$.
$\gamma$ is the PGFCS induced by the triple $(\cE,V,\rho)$.

Our goal in this section is to extend a result on the uniqueness of the representation of
a PGFCS first proved in~\cite[Theorem~1.3]{Fannes1994}. We will mostly follow the alternative
proof of this result given in~\cite[Theorem~2]{Guta2015}. With only minor changes,
the arguments of~\cite{Guta2015} give the following theorem.

\bet\label{thm:equal PGFCS}
Let $\cH_1$ and $\cH_2$ be finite-dimensional Hilbert spaces. For $j\in\{1,2\}$, let
$\Phi_j\in\CPU(\cH_j)$ be irreducible with faithful invariant state $\rho_j$ and Stinespring
dilation $(\cE,V_j)$. Denote by $\gamma_j$ the PGFCS induced by $(\cE,V_j,\rho_j)$.
Then the following statements are equivalent:
\begin{enumerate}[label=(\roman*)]
\item $\gamma_1=\gamma_2$.
\item There exists a unitary $U:\cH_1\rightarrow \cH_2$ such that $U\rho_1=\rho_2U$ and
\begin{equation}\label{eq:uotimesonev2u}
(U\otimes\one_\cE)V_1=\e^{\i\varphi}V_2U
\end{equation}
for some $\varphi\in\RR$.
\end{enumerate}
\eet

In~\cite{Guta2015} this theorem is stated with the stronger assumption that $\Phi_1$ and $\Phi_2$ are primitive. In a private communication, M.~Guta and collaborators informed us they obtained, recently and independently, an improved version of their result matching ours.
In~\cite{Guta2015}, the proof is based on two lemmas. In the generalized setting, the first one,~\cite[Lemma~1]{Guta2015},
translates into the following.

\bel\label{lem:eigenvalue implies unitary}
Under the setting of Theorem~\ref{thm:equal PGFCS}, define $\Phi_{ij}:\cB(\cH_j,\cH_i)\to\cB(\cH_j,\cH_i)$ by
$$
\Phi_{ij}:X\mapsto V_i^*(X\otimes\one_\cE)V_j.
$$
If\, $\Phi_1$ is irreducible, then the three following statements are equivalent:
\begin{enumerate}[label=(\roman*)]
\item $\Phi_{12}$ has an eigenvalue of modulus $1$.
\item $\Phi_{21}$ has an eigenvalue of modulus $1$.
\item There exists a linear isometry $U:\cH_1\to\cH_2$, and $\varphi\in\RR$ such that
Relation~\eqref{eq:uotimesonev2u} holds.
\end{enumerate}
If these statements hold, then $\Phi_{21}(U)=\e^{\i\varphi}U$ and
$\Phi_{12}(U^\ast)=\e^{-\i\varphi}U^\ast$. Moreover, if $\Phi_2$ is irreducible, then $U$ is
unitary and $U\rho_1=\rho_2U$.
\eel

\proof
The equivalence of~(i) and (ii) follows from the fact that $\Phi_{21}(X)=\lambda X$ iff
$\Phi_{12}(X^\ast)=\overline\lambda X^\ast$. Assuming~(iii), one has
$$
\Phi_{21}(U)=V_2^\ast(U\otimes\one_\cE)V_1=\e^{\i\varphi}V_2^\ast V_2U=\e^{\i\varphi}U,
$$
and hence (ii) holds. It remains to show that (ii) implies (iii). Assuming that $\Phi_{21}(X)=\e^{\i\varphi}X$ for
some non-zero $X$, and using the fact that $P_2=V_2V_2^\ast$ is the orthogonal projection onto the range of $V_2$, we get
\begin{align*}
\Phi_{1}(X^\ast X)&=V_1^\ast(X^\ast\otimes\one_\cE)(X\otimes\one_\cE)V_1\\
&\ge V_1^\ast(X^\ast\otimes\one_\cE)P_2(X\otimes\one_\cE)V_1\\
&=\Phi_{21}(X)^\ast\Phi_{21}(X)=X^\ast X,
\end{align*}
and hence
\beq
\Phi_1^n(X^\ast X)\ge X^\ast X
\label{eq:rusalka}
\eeq
for $n\ge1$. Let $\rho$ be a density matrix supported by the
spectral subspace of $X^\ast X$ corresponding to its norm, so that
$\bra\rho,X^\ast X\ket=\|X^\ast X\|$.
Using the fact that $\Phi_1$ is irreducible and that  $\rho_1$ is its unique invariant state, we obtain
$$
\lim_{N\to\infty}\frac1N\sum_{n=1}^N\bra\Phi_1^{\ast n}(\rho),X^\ast X\ket
=\bra\rho_1,X^\ast X\ket.
$$
This is the only change to the proof of~\cite[Lemma~1]{Guta2015}: we use the Ces\`aro mean instead of the limit
of the sequence $(\Phi_1^{\ast n}(\rho))_{n\in\NN}$. This requires only the
irreducibility of $\Phi_1$, whereas primitivity was invoked in~\cite{Guta2015}. Using~\eqref{eq:rusalka}, we have
$\bra\Phi_1^{\ast n}(\rho),X^\ast X\ket=\bra\rho,\Phi_1^n(X^\ast X)\ket\ge\bra\rho,X^\ast X\ket$,
and we deduce
$$
\bra\rho_1,X^\ast X\ket\geq\bra\rho,X^\ast X\ket=\|X^\ast X\|.
$$
Setting $D=\|X^\ast X\|\one_{\cH_1}-X^\ast X$, we have $D\ge0$ while $\bra\rho_1,D\ket\le0$.
Since $\rho_1>0$, it follows that $D=0$, \ie that $U=\|X\|^{-1}X$ is an isometry,
$U^\ast U=\one_{\cH_1}$, such that
$$
V_2^\ast(U\otimes\one_\cE)V_1=\Phi_{21}(U)=\e^{\i\varphi}U,
$$
and hence, after left multiplication with $V_2$,
\beq
P_2(U\otimes\one_\cE)V_1=\e^{\i\varphi}V_2U.
\label{christie}
\eeq
Setting $Y=(\one_{\cH_2}\otimes\one_\cE-P_2)(U\otimes\one_\cE)V_1$, we note that
$$
Y^\ast Y=V_1^\ast(U^\ast\otimes\one_\cE)(U\otimes\one_\cE)V_1-U^\ast V_2^\ast V_2U=V_1^\ast V_1-U^\ast U=0,
$$
so that~\eqref{eq:uotimesonev2u} now follows from~\eqref{christie}.

Finally, we note that Relation~\eqref{eq:uotimesonev2u} implies
$U^\ast\Phi_2(X)U=\Phi_1(U^\ast XU)$ for all $X\in\cB(\cH_2)$. By duality,
$$
U\rho_1U^\ast=U\Phi_1^\ast(\rho_1)U^\ast=\Phi_2^\ast(U\rho_1U^\ast).
$$
Thus, if $\Phi_2$ is irreducible, we can conclude that $U\rho_1U^\ast=\rho_2>0$,
which implies $\Ran U=\cH_2$ and that $U$ is unitary.\QED

We replace the second result~\cite[Lemma~2]{Guta2015} involved in the proof of~\cite[Theorem~2]{Guta2015}
with the following self-contained lemma which provides some form of decoherence without assuming
any irreducibility of the involved quantum channel.

\bel\label{lem:decoherent dilations}
Let $\cH$ be a finite-dimensional Hilbert space and $\rho$ a density matrix on $\cH$. Let $(\cE^{(n)})_{n\in \NN}$
be a sequence of Hilbert spaces. For $n\in\NN$, let $V^{(n)}:\cH\to\cH\otimes\cE^{(n)}$ be a linear isometry, and
denote by $\Psi^{(n)}\in\CPU(\cH)$ the associated quantum channel
$X\mapsto V^{(n)\ast}(X\otimes \one_{\cE^{(n)}})V^{(n)}$.

If the orthogonal projections $P_j\in\BH$, $j\in\{1,2\}$, are such that
\beq
\inf_{n\in\NN}\bra\rho,\Psi^{(n)}(P_j)\ket>0,
\label{eq:inf_probab_PQ}
\eeq
then
$$
\gamma_j^{(n)}=\frac{\tr_\cH((P_j\otimes\one_{\cE^{(n)}})V^{(n)}\rho V^{(n)\ast}
(P_j\otimes\one_{\cE^{(n)}}))}{\bra\rho,\Psi^{(n)}(P_j)\ket}
$$
are density matrices on $\cE^{(n)}$ satisfying
\beq
\inf_{n\in \NN}\|\gamma_j^{(n)}\|_\HS>0.
\label{eq:infnsigmapq}
\eeq
Moreover, if for any $X\in\BH$,
\beq
\lim_{n\to\infty}\Psi^{(n)}(P_1XP_2)=0,
\label{eq:poq0}
\eeq
then
\beq
\lim_{n\to\infty}\bra\gamma_1^{(n)},\gamma_2^{(n)}\ket_\HS=0.
\label{mars-1}
\eeq
\eel

\ber Since both
states have bounded rank, \eqref{mars-1} gives that  $\lim_n\|\gamma_1^{(n)}-\gamma_2^{(n)}\|_1=2$.
\eer

\proof
Obviously, $\gamma_j^{(n)}\in\cB_+(\cE^{(n)})$, and the cyclicity of the trace gives
$$
\tr((P_j\otimes\one_{\cE^{(n)}})V^{(n)}\rho V^{(n)\ast}(P_j\otimes\one_{\cE^{(n)}}))
=\tr(\rho V^{(n)\ast}(P_j\otimes\one_{\cE^{(n)}})V^{(n)})=\bra\rho,\Psi^{(n)}(P_j)\ket,
$$
so that $\tr(\gamma_j^{(n)})=1$. Moreover, since $\cH$ is finite dimensional, $\gamma_j^{(n)}$ has finite rank $r\le\dim(\cH)^2$.
Denoting by $(\kappa_k)$ its non-vanishing eigenvalues, it follows from the Cauchy--Schwarz inequality that
$$
1=\tr(\gamma_j^{(n)})=\sum_{k=1}^r\kappa_k
\le\left(\sum_{k=1}^r1\right)^{1/2}\left(\sum_{k=1}^r\kappa_k^2\right)^{1/2}
\le\dim(\cH)\,\|\gamma_j^{(n)}\|_\HS,
$$
and~\eqref{eq:infnsigmapq} follows.

Denoting by  $(e_k)$ and $(f_l)$ orthonormal bases of $\Ran P_1$ and $\Ran P_2$, an elementary calculation leads to
\begin{align*}
\bra\gamma_1^{(n)},\gamma_2^{(n)}\ket_\HS
=\sum_{k,l}\frac{\|\rho^{1/2}\Psi^{(n)}(e_k\otimes f_l^\ast)\rho^{1/2}\|_\HS^2}
{\bra\rho,\Psi^{(n)}(P_1)\ket\bra\rho,\Psi^{(n)}(P_2)\ket},
\end{align*}
and Relations~\eqref{eq:inf_probab_PQ} and~\eqref{eq:poq0} yield the last assertion of the lemma.\QED

\noindent{\bf Proof of Theorem~\ref{thm:equal PGFCS}}.
The argument follows~\cite{Guta2015}. We give details for the reader's convenience. 

\noindent{$(ii)\Rightarrow(i)$} Setting $u:\cB(\cH_1)\ni X\mapsto u(X)=UXU^\ast\in\cB(\cH_2)$, one
easily deduces from~(ii) that
$$
\fS_{V_1,A}\circ u^\ast=u^\ast\circ\fS_{V_2,A},
$$
and $u(\rho_1)=\rho_2$. Since $u^\ast(\one_{\cH_2})=\one_{\cH_1}$, one has
\begin{align*}
\gamma_1(A_1\otimes\cdots\otimes A_n)&=\bra\rho_1,\fS_{V_1,A_1}\cdots\fS_{V_1,A_n}\one_{\cH_1}\ket_\HS\\
&=\bra\rho_1,\fS_{V_1,A_1}\cdots\fS_{V_1,A_n}u^\ast\one_{\cH_2}\ket_\HS\\
&=\bra\rho_1,\fS_{V_1,A_1}\cdots u^\ast\fS_{V_2,A_n}\one_{\cH_2}\ket_\HS\\
&=\bra\rho_1,u^\ast\fS_{V_2,A_1}\cdots\fS_{V_2,A_n}\one_{\cH_2}\ket_\HS\\
&=\bra u\rho_1,\fS_{V_2,A_1}\cdots\fS_{V_2,A_n}\one_{\cH_2}\ket_\HS\\
&=\bra\rho_2,\fS_{V_2,A_1}\cdots\fS_{V_2,A_n}\one_{\cH_2}\ket_\HS\\
&=\gamma_2(A_1\otimes\cdots\otimes A_n).
\end{align*}

\noindent{$(i)\Rightarrow(ii)$} Let $\cH=\cH_1\oplus\cH_2$ and denote by $P_1,P_2\in\BH$ the
orthogonal projections onto the subspaces $\cH_1\oplus\{0\}$ and $\{0\}\oplus\cH_2$. Setting
$V=V_1\oplus V_2:\cH\to\cH\otimes\cE$, define $\Psi\in\CPU(\cH)$ by
$$
\Psi:X\mapsto V^\ast(X\otimes\one_\cE)V.
$$
With the notations of Lemma~\ref{lem:eigenvalue implies unitary}, this reads
$$
\Psi:\begin{bmatrix}
X_{11}&X_{12}\\
X_{21}&X_{22}
\end{bmatrix}
\mapsto
\begin{bmatrix}
\Phi_1(X_{11})&\Phi_{12}(X_{12})\\
\Phi_{21}(X_{21})&\Phi_2(X_{22})
\end{bmatrix},
$$
and hence the four subspaces $P_i\BH P_j\subset\BH$ are $\Psi$-invariant.

For $n\in\NN$, let $\cE^{(n)}=\cE^{\otimes n}$, and note that
$$
V^{(n)}=(V\otimes\one_{\cE^{(n-1)}})\cdots(V\otimes\one_{\cE^{(2)}})(V\otimes\one_{\cE^{(1)}})V
$$
defines an isometry $V^{(n)}:\cH\to\cH\otimes\cE^{(n)}$. Thus, we can define $\Psi^{(n)}\in\CPU(\cH)$
as in Lemma~\ref{lem:decoherent dilations}, and $\fS_{V^{(n)},O}$ with $O\in\cB(\cE^{(n)})$ as in
Definition~\ref{def:PsiV}(i).
For $A_1,\ldots,A_n\in\cB(\cE)$, we have
$$
\fS_{V^{(n)},A_n\otimes\cdots\otimes A_1}=\fS_{V^{(n-1)},A_{n-1}\otimes\cdots\otimes A_1}\circ\fS_{V,A_n},
$$
and hence
$$
\fS_{V^{(n)},A_n\otimes\cdots\otimes A_1}=\fS_{V,A_1}\circ\cdots\circ\fS_{V,A_n}.
$$
In particular, with $A_1=\cdots=A_n=\one_{\cE}$,
$$
\Psi^{(n)}=\Psi^n.
$$
Setting $\rho=\frac12(\rho_1\oplus\rho_2)$, we get
$$
\bra\rho,\Psi^{(n)}(P_j)\ket=\bra\rho,\Psi^n(P_j)\ket=
\tfrac12\bra\rho_j,\Phi_j^n(\one_{\cH_j})\ket=\tfrac12,
$$
so that Assumption~\eqref{eq:inf_probab_PQ} in Lemma~\ref{lem:decoherent dilations} is satisfied.
Defining the density matrices $\gamma_j^{(n)}$ as in this lemma, we get,
for $A_1,\ldots,A_n\in\cB(\cE)$,
\begin{align*}
\bra\gamma_j^{(n)},A_n\otimes\cdots\otimes A_1\ket_\HS
&=2\,\tr\left((P_j\otimes\one_{\cE^{(n)}})V^{(n)}\rho V^{(n)\ast}(P_j\otimes\one_{\cE^{(n)}})
(A_n\otimes\cdots\otimes A_1)\right)\\
&=2\,\bra\rho,\fS_{V^{(n)},A_n\otimes\cdots\otimes A_1}(P_j)\ket
=2\,\bra\rho,\fS_{V,A_1}\cdots\fS_{V,A_n}P_j\ket_\HS.
\end{align*}
Since
$$
\fS_{V,A}:\begin{bmatrix}
X_{11}&X_{12}\\
X_{21}&X_{22}
\end{bmatrix}
\mapsto
\begin{bmatrix}
V_1^\ast(X_{11}\otimes A)V_1&V_1^\ast(X_{12}\otimes A)V_2\\
V_2^\ast(X_{21}\otimes A)V_1&V_2^\ast(X_{22}\otimes A)V_2
\end{bmatrix},
$$
we conclude that
$$
\bra\gamma_j^{(n)},A_n\otimes\cdots\otimes A_1\ket_\HS
=\gamma_j(A_1\otimes\cdots\otimes A_n\otimes\one).
$$
From $\Psi\in\CPU(\cH)$ we infer that all its eigenvalues lie in the closed unit disc. Hence, the same holds
for $\Phi_{12}$, its restriction to the invariant subspace $P_1\BH P_2$. We proceed to show that $\Phi_{12}$ has an eigenvalue of modulus $1$. Suppose it is not so, namely that all the eigenvalues of $\Phi_{12}$
have modulus strictly less than $1$. Then, for any $X\in\BH$,
$$
\lim_{n\to\infty}\Psi^n(P_1XP_2)=0,
$$
and Lemma~\ref{lem:decoherent dilations} implies that $\lim_{n}\bra\gamma_1^{(n)},\gamma_2^{(n)}\ket_\HS=0$.
But by~\eqref{eq:infnsigmapq}, we have $\bra\gamma_1^{(n)},\gamma_1^{(n)}\ket_\HS\ge\delta$ for
some $\delta>0$. So $\gamma_1^{(n)}\neq\gamma_2^{(n)}$ for large enough $n$, which contradicts the assumption $\gamma_1=\gamma_2$. Therefore, $\Phi_{12}$ has an eigenvalue of modulus $1$, and~(ii) follows from
Lemma~\ref{lem:eigenvalue implies unitary}.\QED

\subsection{Proof of Theorem~\ref{main-instrument-QDB}}
\label{sec:TRI implies QDB}

We start by noticing that, the state $\rho$ being faithful, Assumption~\ER{} is satisfied as a consequence of Lemma~\ref{lem:proof ER}.

\noindent{$(i)\Rightarrow(ii)$} Invoking successively~\eqref{eq:hatJform},~\IQDB, and \eqref{eq:Pdef}, we have
\begin{align*}
\wP([A_1,\ldots,A_n])
&=\bra\one,\wJ(A_1)\cdots\wJ(A_n)\one\ket_\KMS\\
&=\bra\one,\cJ(A_1)\cdots\cJ(A_n)\one\ket_\KMS
=\PP([A_1,\ldots,A_n]),
\end{align*}
for any $A_1,\ldots,A_n\in\cA$, and hence $\wP=\PP$.

It remains to prove $\theta$ is unitarily (resp. anti-unitarily) implementable. Let $(\cE,V)$ be a Stinespring dilation of $\Phi$ such that $\cE_V=\cE$. Since~\IQDB{} implies~\QDB{}, $(\cE,V)$ is also a dilation of $\widehat{\Phi}$. Let $M$ be the $(\cJ,\cE,V)$-POVM of Proposition~\ref{prop:dilation}(ii). Then, Lemma~\ref{lem:C involution} implies there exists $S:\cE\to\cE$ unitary (resp. anti-unitary) such that
$$\wJ(A)(X)=V^*(X\otimes S^*M\circ\theta(A)S)V.$$
Then \IQDB{} implies that
$$\cJ(A)(X)=V^*(X\otimes M(A))V=V^*(X\otimes S^*M\circ\theta(A)S)V=\wJ(A)(X).$$
Since $\cE=\cE_V$ by hypothesis, it follows that $M\circ\theta(A)=S^*M(A)S$ and $\theta$ is unitarily (resp. anti-unitarily) implementable setting $\Theta=S$.

\noindent{$(ii)\Rightarrow(iii)$} The reversal $\theta$ is implementable by assumption. Proposition~\ref{prop:conv_ep}(i) and time-reversal invariance $\wP=\PP$ give that
$$
\ep(\cJ,\rho,\theta)=\ep(\PP,\theta)=
\lim_{n\to\infty}\tfrac1n\Ep(\PP_n,\theta)
=\lim_{n\to\infty}\tfrac1n\Ent(\PP_n\,|\,\wP_n)=0.
$$

\noindent{$(iii)\Rightarrow(ii)$}
The channel $\Phi$ being irreducible,
the measure $\PP$ is $\phi$-ergodic by Proposition~\ref{prop:prop channel}(v).
Thus, Proposition~\ref{prop:conv_ep}(ii) allows us to conclude that $\wP=\PP$.

\noindent{$(ii)\Rightarrow(i)$} Assuming, in addition, that the associated instrument $\cJ$
is informationally complete, we have to show that~\IQDB{} holds.

Let $(\cE,V)$ be a Stinespring dilation of $\Phi=\cJ(\bA)$ and $M$ an informationally complete
 $(\cJ,\cE,V)$-POVM such that
\beq
\cJ(A)=\fS_{V,M(A)}
\label{eq:def instrument forward}
\eeq
and such that there is a unitary/anti-unitary operator $\Theta:\cE\to\cE$ implementing $\theta$. By assumption, such a choice is always possible.

Let $J$ be an arbitrary anti-unitary/unitary $(\Phi,\rho)$-admissible map and define the
time-reversed instrument $\wJ$ by Relation~\eqref{Crooks}.

Let $(e_l)_{l\in \Lambda}$ be an (at most countable) orthonormal basis of $\cE$. For each $l\in \Lambda$, let $V_l: \cH\to\cH$ be defined by $V_l=(\one_\cH\otimes e_l^*)V$. Then, $\widetilde{V}=\sum_{l\in \Lambda} j^{-1}(\rho^{\frac12}V_l^*\rho^{-\frac12})\otimes e_l$ is a map from $\cH$ to $\cH\otimes\cE$ that is a Stinespring dilation of $\widehat{\Phi}$.

Lemma~\ref{lem:def C}
yields a linear/anti-linear partial isometry
$\widetilde{U}:\cE_{\widetilde{V}}\to\cE_V$ such that
$$
\wJ(A)=\widehat{\fS}_{V,M\circ\theta(A)}=
\fS_{\widetilde{V},\widetilde{U}^\ast M\circ\theta(A)\widetilde{U}}
=\fS_{\widetilde{V},(\Theta\widetilde{U})^\ast M(A)\Theta\widetilde{U}}.
$$
Since $\widetilde{U}$ and $\Theta$ are both linear/anti-linear, $S=\Theta\widetilde{U}$ is always linear.
Thus, setting $\widehat{V}=(\one_\cH\otimes S)\widetilde{V}$, we obtain a Stinespring
dilation $(\cE,\widehat{V})$ of $\widehat{\Phi}$ such that
\beq
\wJ(A)=\fS_{\widehat{V},M(A)}.
\label{eq:def instrument backward}
\eeq
For $n\in\NN$, we set
$$
M^{(n)}(A_1\times\cdots\times A_n)=M(A_1)\otimes\cdots\otimes M(A_n),
$$
and denote by $\gamma$ and $\widehat{\gamma}$ the PGFCS over $\cB(\cE)^{\otimes\NN}$ induced by
$(\cE,V,\rho)$ and $(\cE,\widehat{V},\rho)$. Relations~\eqref{eq:Pdef}-\eqref{eq:hatJform} and the definition of $\gamma^{(n)}$ in~\eqref{eq:pgfcsDef} combined with
Relations~\eqref{eq:def instrument forward}-\eqref{eq:def instrument backward} give
\begin{align*}
\gamma^{(n)}\circ M^{(n)}(A_1\times\cdots\times A_n)
&=\bra\one,\cJ(A_1)\cdots\cJ(A_n)\one\ket_\KMS=\PP([A_1,\ldots,A_n]);\\[4pt]
\widehat{\gamma}^{(n)}\circ M^{(n)}(A_1\times\cdots\times A_n)
&=\bra\one,\wJ(A_1)\cdots\wJ(A_n)\one\ket_\KMS=\wP([A_1,\ldots,A_n]).
\end{align*}
By Lemma~\ref{lem:guillaumetell}, $M^{(n)}$ is informationally complete, and 
the identity $\PP=\wP$ allows us to conclude that $\gamma^{(n)}=\widehat{\gamma}^{(n)}$
for all $n\in\NN$, and hence that $\gamma=\widehat{\gamma}$.

Since $\Phi$ is assumed to be irreducible, so is $\widehat{\Phi}$.
Thus, Theorem~\ref{thm:equal PGFCS} implies that there exists a unitary
$U\in\BH$ and $\varphi\in\RR$ such that $U\rho=\rho U$ and
$$
\e^{\i\varphi}VU=(U\otimes\one_\cE)\widehat{V}.
$$
Taking~\eqref{eq:def instrument backward} into account, it follows that for any $A\in\cA$
and $X\in\BH$,
$$
\cJ(A)(X)=V^\ast(X\otimes M(A))V=U\widehat{V}^\ast(U^\ast XU\otimes M(A))\widehat{V}U^\ast
=u\circ\wJ(A)\circ u^{-1}(X),
$$
where $u(X)=UXU^\ast$ satisfies $u(\rho)=\rho$ and  is therefore unitary
w.r.t.\;the KMS inner product. Setting $G=JU^\ast$ and $g=j\circ u^{-1}$,
we have
\beq
\cJ(A)=g^{-1}\circ\cJ(\theta(A))^\rho\circ g=\hat{\cJ}(A),
\label{eq:flu}
\eeq
and it remains to show that $g$ is $(\Phi,\rho)$-admissible.

As already mentioned, $u(\rho)=\rho$, so $g(\rho)=\rho$, and hence $g$
is $\rho$-anti-unitary/unitary. Taking the KMS-dual of Relation~\eqref{eq:flu} yields
$$
\cJ(\theta(A))^\rho=g^{-1}\circ\cJ(A)\circ g,
$$
which, inserted into~\eqref{eq:flu}, gives $g^2\circ\cJ(A)=\cJ(A)\circ g^2$,
\ie with $\Psi(Y)=V^\ast YV$,
$$
g^2\circ\Psi(X\otimes M(A))=\Psi\circ(g^2\otimes\id_{\cB(\cE)})(X\otimes M(A)).
$$
Invoking Lemma~\ref{lem:IC}(ii) and weak*-continuity, we derive
$$
g^2\circ\Psi=\Psi\circ(g^2\otimes\id_{\cB(\cE)}),
$$
and so
$$
(g^2\circ\Psi)(Y)
=(G^2V^\ast)Y(G^2V^\ast)^\ast
=(V^\ast(G^2\otimes\one_\cE))Y
(V^\ast(G^2\otimes\one_\cE))^\ast.
$$
The two expressions on the right are minimal Stinespring representations
of $g^2\circ\Psi$ with the identity representation of
$\cB(\cH\otimes\cE)$. Since
$\cB(\cH\otimes\cE)'=\CC\one_{\cH\otimes\cE}$, Stinespring uniqueness
gives
$$
V^\ast(G^2\otimes\one_\cE)=\e^{\i\varphi}G^2V^\ast
$$
for some $\varphi\in\RR$, so that
$$
\Phi(G^2)=V^\ast(G^2\otimes\one_\cE)V=\e^{\i\varphi}G^2V^\ast V=\e^{\i\varphi}G^2.
$$
Thus $G$ and $g$ satisfy the two relations~\eqref{eq:Phi-rho-adm} ensuring their $(\Phi,\rho)$-admissibility,
and Relation~\eqref{eq:flu} shows that $\cJ$ satisfies the \IQDB{} condition.
\QED



\printbibliography[
  heading=bibintoc,
  title={References}
]
\end{document}